\documentclass[aps,prd,showpacs,notitlepage,nofootinbib,superscriptaddress,floatfix,showkeys,twocolumn]{revtex4-2}
\usepackage{blindtext}
\usepackage{hyperref}

\usepackage{amsmath,amssymb}
\usepackage{float}
\usepackage{microtype}

\usepackage{graphicx}
\usepackage{bm}
\usepackage{latexsym}
\usepackage{epsfig}
\usepackage{psfrag}
\usepackage[dvipsnames]{xcolor}
\usepackage{subfigure}
\usepackage{graphicx}

\usepackage{amssymb}
\usepackage{amsmath}
\usepackage{bm}
\usepackage{latexsym}
\usepackage{epsfig}
\usepackage{psfrag}
\usepackage[normalem]{ulem}
\usepackage{textcomp}
\usepackage{color}
\usepackage{pstricks}
\usepackage[utf8]{inputenc}
\usepackage{comment}
\usepackage{orcidlink}

\def\beq{\begin{equation}}
\def\eeq{\end{equation}}
\def\bea{\begin{eqnarray}}
\def\eea{\end{eqnarray}}
\def\be{\begin{equation}}
\def\ee{\end{equation}}

\def\bse{\begin{subequations}}
\def\ese{\end{subequations}}

\def\cF{\mathcal{F}}

\def\ee{\eta_{\rm e}}

\def\Mpl{M_{_{\mathrm{Pl}}}}
\def\f{\frac}
\def\l{\left}
\def\r{\right}
\def\d{\mathrm{d}}
\def\e{{\mathrm{e}}}
\def\fpbh{f_{_{\rm PBH}}}
\def\nn{\nonumber}

\def\tin{t_\mathrm{in}}
\def\tre{t_{\mathrm {re}}}

\def\tq{t_{\mathrm {q}}}
\def\tbh{t_{_{\mathrm {BH}}}}
\def\tac{t_{\mathrm {acc}}}
\def\tev{t_{\mathrm {ev}}^0}
\def\tevk{t_{\mathrm {ev}}^{\kappa}}
\def\tbbn{t_{_{\mathrm {BBN}}}}
\def\Tbh{T_{_{\mathrm {BH}}}}
\def\Tev{T_{\mathrm {ev}}}
\def\Tre{T_{\mathrm {re}}}

\def\mac{M_{\mathrm {acc}}}
\def\Min{M_\mathrm{in}}
\def\hin{H_\mathrm{in}}

\def\rhobh{\rho_{_\mathrm{BH}}}
\def\rhor{\rho_{_\mathrm{R}}}
\def\rhow{\rho_{w}}
\def\betacr{\beta_{\mathrm{cr}}}
\def\nbh{n_{_\mathrm{BH}}}
\def\lc{\lambda_{\mathrm{c}}}
\def\lcr{\lambda_{\mathrm{c}}^{\mathrm{R}}}
\def\lcnr{\lambda_{\mathrm{c}}^{\mathrm{NR}}}
\def\dneff{\Delta N_{\mathrm{eff}}}

\def\ain{a_\mathrm{in}}
\def\are{a_\mathrm{re}}
\def\aeq{a_\mathrm{eq}}
\def\aev{a_\mathrm{ev}}

\def\obh{\Omega_{_\mathrm{PBH}}}

\begin{document}

\title{Relativistic accretion and burdened primordial black holes}
\author{Suvashis Maity\,\orcidlink{0009-0004-9960-7802}}
\email{E-mail: saamaity@gmail.com}
\affiliation{Indian Institute of Science Education and Research, Pune 411008, India}

\begin{abstract}
We examine the joint effects of relativistic accretion and memory-burdened evaporation on the evolution of primordial black holes (PBHs). The memory burden effect, which delays the evaporation by inducing a backreaction and making the evaporation rate scale as an inverse power law of the PBH entropy, opens up a new window that allows PBHs with $M \lesssim 10^{15}~\mathrm{g}$ to survive until the present epoch. Meanwhile, accretion increases the mass of PBHs, thereby enhancing their chances of survival for a given initial mass.
We consider two main scenarios: one where PBHs evaporate completely before big bang nucleosynthesis, and another where PBHs persist until today. In the case of evaporation, we analyze the emission of dark matter (DM) and dark radiation (DR) during the process of evaporation. Conversely, in the other case, the surviving PBHs themselves can contribute as DM. We further investigate how relativistic and nonrelativistic accretion, together with memory-burdened evaporation, impact the parameter space of the emitted DM, the abundance of stable PBHs as DM, and the contribution of DR to the effective number of relativistic degrees of freedom, $\Delta N_{\mathrm{eff}}$.
\end{abstract}
\maketitle


\section{Introduction \label{sec: Introduction}}

Primordial black holes (PBHs) have attracted widespread interest over the past few decades~\cite{Carr:1974nx}.
They are thought to form from the collapse of overdense regions in the early Universe.
Due to their cold and noninteracting nature, PBHs are also considered a compelling candidate for dark matter (DM)~\cite{Hawking:1971ei,Carr:1974nx,Ivanov:1994pa,Bartolo:2018evs,Cai:2018dig,Carr:2020xqk,Jedamzik:2020ypm,Jedamzik:2020omx,Green:2020jor,Villanueva-Domingo:2021spv,Carr:2021bzv}.
Recent detections of binary black hole mergers by the LIGO-Virgo collaboration have further revived interest in PBHs, as they could potentially contribute to the observed merger rate~\cite{LIGOScientific:2016aoc,LIGOScientific:2016dsl,LIGOScientific:2016wyt,
LIGOScientific:2017bnn,LIGOScientific:2017ycc,LIGOScientific:2017vox}.
Moreover, there have been numerous recent efforts to constrain the fraction of DM that can be in the form of PBHs by analyzing scalar-induced secondary gravitational waves (GWs) probed by pulsar timing arrays~\cite{NANOGrav:2023gor, NANOGrav:2023hde,EPTA:2023sfo,EPTA:2023fyk,Zic:2023gta, Reardon:2023gzh,Xu:2023wog}.   

PBHs are also metastable objects that can evaporate over time, emitting various particles through the process of evaporation~\cite{Hawking:1974rv, Hawking:1975vcx}.
During evaporation, PBHs can emit standard model (SM) particles~\cite{Sandick:2021gew,RiajulHaque:2023cqe,Maity:2024cpq} (which may potentially reheat the Universe), contribute to the relic abundance of DM~\cite{Green:1999yh,Khlopov:2008qy,Belotsky:2014kca,Gondolo:2020uqv,Cheek:2021odj,Cheek:2021cfe,Bernal:2022oha,Khlopov:2024nqp}, generate dark radiation (DR)~\cite{Hooper:2019gtx,Masina:2020xhk,Arbey:2021ysg,Barman:2024iht} or produce high-frequency GWs~\cite{Dong:2015yjs,Sugiyama:2020roc,Inomata:2020lmk,Domenech:2021wkk,Ireland:2023avg}, thereby influencing several cosmological observables.
In a semiclassical framework, PBHs undergo Hawking evaporation, leading to stringent constraints: PBHs with masses $M < 10^{15}~\mathrm{g}$ would have largely evaporated by today, while those with masses $M > 10^{9}~\mathrm{g}$ would have evaporated after about one second, thus impacting BBN and being constrained by the abundance of light elements.

Recent studies have revealed that quantum effects play a pivotal role in the evaporation of PBHs at later stages. As a PBH loses mass via Hawking radiation, its associated entropy decreases. Since black hole entropy encodes the number of microstates, this loss leads to a reduction in the information-storing capacity of the system. However, if evaporation of PBHs is unitary, as demanded by quantum mechanics, this loss of information must be accounted for. The resulting backreaction on the process of evaporation, arising from the burden of retaining information is known as the memory burden effect~\cite{Dvali:2012en,Dvali:2020wft,Michel:2023ydf,Wang:2023wsm}. This effect can significantly slow down or even stabilize the evaporation, effectively allowing PBHs to persist longer than predicted by standard Hawking evaporation.

One of the most striking consequences is that PBHs with masses below $10^{15}$ g, which would otherwise have evaporated completely by the present epoch, may still survive. This opens up a new mass window in the allowed parameter space for the fraction of DM contributed by PBHs, $\fpbh$
~\cite{Alexandre:2024nuo,Thoss:2024hsr,Haque:2024eyh,Chaudhuri:2025asm}. Moreover, the memory burden effect~\cite{Dvali:2018xpy,Dvali:2018ytn} has been shown to impact a variety of cosmological observables and theoretical processes. These include potential signatures in the stochastic GW background~\cite{Moursy:2024hll,Barman:2024iht,Bhaumik:2024qzd,Kohri:2024qpd,Jiang:2024aju,Barker:2024mpz,Gross:2024wkl,Athron:2024fcj}, implications for DM~\cite{Haque:2024eyh,Saha:2024ies,Barker:2024mpz,Loc:2024qbz,Borah:2024bcr,Montefalcone:2025akm,Dvali:2025ktz,Takeshita:2025mhy}, baryogenesis mechanisms such as leptogenesis~\cite{DeRomeri:2024zqs,Chianese:2024rsn,Calabrese:2025sfh}, and other phenomena that remain active areas of investigation~\cite{Khlopov:2020vpx,Dvali:2024hsb,Shallue:2024hqe,Zantedeschi:2024ram,Basumatary:2024uwo,Barman:2024kfj,Bandyopadhyay:2025ast,Boccia:2025hpm,Chianese:2025wrk,Tan:2025vxp,Dondarini:2025ktz}.

Other than evaporation, PBHs can also accrete matter from their surroundings, potentially increasing their mass. For PBHs modeled as Newtonian point masses embedded in an infinite homogeneous gas cloud, Bondi~\cite{Bondi:1952ni} (for related discussions, see~\cite{Harada:2004pf,Harada:2004pe,Mack:2006gz,Lora-Clavijo:2013aya,Dihingia:2018tlr,DeLuca:2020fpg,DeLuca:2021pls,Yang:2021agk,Yang:2022puh,Yuan:2023bvh,Zhang:2023rnp,Jangra:2024sif}) first analyzed the spherically symmetric, steady-state accretion flow. However, when the velocity of the infalling matter approaches relativistic speeds or the background temperature becomes high, the accretion dynamics require relativistic corrections to the classical Bondi-Hoyle formalism~\cite{Aguayo-Ortiz:2021jzv,Das:2025vts}. In such a regime, the rate of accretion becomes highly sensitive to the mass of PBHs, the energy density of the surrounding medium, and its equation of state (EoS). Recent studies have shown that relativistic accretion can substantially enhance the growth of the mass of PBHs, particularly when they form during early, dense phases~\cite{Das:2025vts}. These effects can modify the mass of PBHs, affect their survival probability, and lead to important implications for cosmological constraints.

In this work, we have investigated the relativistic accretion of PBHs with a lower mass that would otherwise have completely evaporated before the onset of BBN via Hawking radiation. 
We examine how relativistic accretion influences the process of evaporation, particularly when combined with the effects of memory burden. The accretion can further prolong the lifetime of such PBHs, potentially allowing them to survive past the epoch of BBN. We also study the production of DM from the evaporation of PBHs and examine how both relativistic accretion and memory burden modify the allowed parameter space for DM emitted. Furthermore, for PBHs that may still exist today due to memory burden effects, we apply existing constraints on $\fpbh$ in the newly opened mass window $(<10^{15}~\mathrm{g})$~\cite{Alexandre:2024nuo,Thoss:2024hsr,Haque:2024eyh,Chaudhuri:2025asm}. Using these, we derive bounds on the initial abundance of PBHs and demonstrate how these constraints differ from scenarios that neglect accretion. Further, we consider the emission of DR from the evaporation of PBHs and discuss the contribution of DR to the $\dneff$, especially in the light of relativistic accretion and burdened evaporation.

This paper is organisedd as follows.
In Sec.~\ref{sec: mass-evol}, we discuss the evolution of the mass of PBHs, taking into account both relativistic accretion and burdened evaporation. 
In Sec.~\ref{sec:dm-evap}, we examine the DM produced from the evaporation of PBHs and explore how accretion and evaporation affect the parameter space of the emitted DM.
Further, in Sec.~\ref{sec:dm-stable}, we discuss the stable PBHs in the new mass window arising due to the memory burden, and analyze the impact of accretion on this scenario. 
In Sec.~\ref{sec:DR-evap}, we investigate the contribution to $\dneff$ from DR generated through evaporation. 
Finally, in Sec.~\ref{sec:conclusion}, we conclude with a brief summary.

We shall now make a few clarifying remarks on the conventions and notations that we shall adopt in this work. 
We shall work with natural units such that $\hbar=c=1$, and the reduced 
Planck mass is $\Mpl=\l(8\,\pi\, G\r)^{-1/2} \simeq 2.4 \times10^{18}\,
\mathrm{GeV}$.
The signature of the metric is followed~$(-,+,+,+)$ and the background is spatially flat
Friedmann-Lema\^itre-Robertson-Walker~(FLRW).
An overdot and an overprime shall denote differentiation with 
respect to the cosmic time~$t$ and the conformal time~$\eta$, respectively. Derivatives with respect to any other quantity will be denoted as subscript by the quantity.

\section{Evolution of the mass of PBHs \label{sec: mass-evol}}

We begin by discussing the evolution of the mass of a PBH under the combined effects of accretion and evaporation. PBHs are assumed to form from the collapse of overdense regions in the early Universe, with the mass at the time of formation related to the Hubble parameter at that epoch by
 $\Min=4\pi\gamma\Mpl^2\hin^{-1}$, where 
$\gamma=w^{3/2}$ is the collapse efficiency, which quantifies the fraction of the mass enclosed within the Hubble radius that collapses to form PBHs. For a background with a general EoS
$w$, the Hubble parameter evolves with cosmic time as $H=2/[3(1+w)t]$. 
Substituting this expression into the equation for $\Min$, we obtain the relation between the time of the formation of PBHs $\tin$ and the initial mass of PBHs to be
\begin{align}
    \tin=\f{\Min}{6\pi\gamma(1+w)\Mpl^2}.
\end{align}
Once the value of $\hin$ is determined, we can calculate the corresponding $\rho_w^{\rm in}=3\Mpl^2\hin^2$. The initial energy density of PBHs is then given by $\rhobh^{\rm in}=\beta \rho_w^{\rm in}$. The parameter $\beta$ represents the initial abundance of PBHs and depends on several factors, including the nature of the inflationary scalar power spectrum. However, calculating $\beta$ requires a detailed analysis beyond the scope of this work. Therefore, we treat $\beta$ as a free parameter.
Other than the mass, two important quantities associated with PBHs are their temperature $\Tbh$ and entropy $S$, both of which are functions of the PBH mass as 
\begin{align}
    \Tbh=\f{\Mpl^2}{M}, \quad S=\f{1}{2}\l(\f{M}{\Mpl}\r)^2.
    \label{eq:temp-entrop}
\end{align}
The above relations in Eq.~\eqref{eq:temp-entrop}, indicate that a PBH with lower mass will have higher energy but lower entropy. 
The energy density of PBHs, $\rhobh$ is related to the number density $\nbh$ as $\rhobh=\nbh\times M$. 
While the cosmic expansion causes the number density to dilute as $\nbh\sim a^{-3}$, the mass of individual PBHs is not conserved, as it evolves due to the competing effects of accretion and evaporation. In what follows, we will discuss in detail how these two processes influence the evolution of mass over time.
\begin{figure*}
    \centering
    \includegraphics[width=0.49\linewidth]{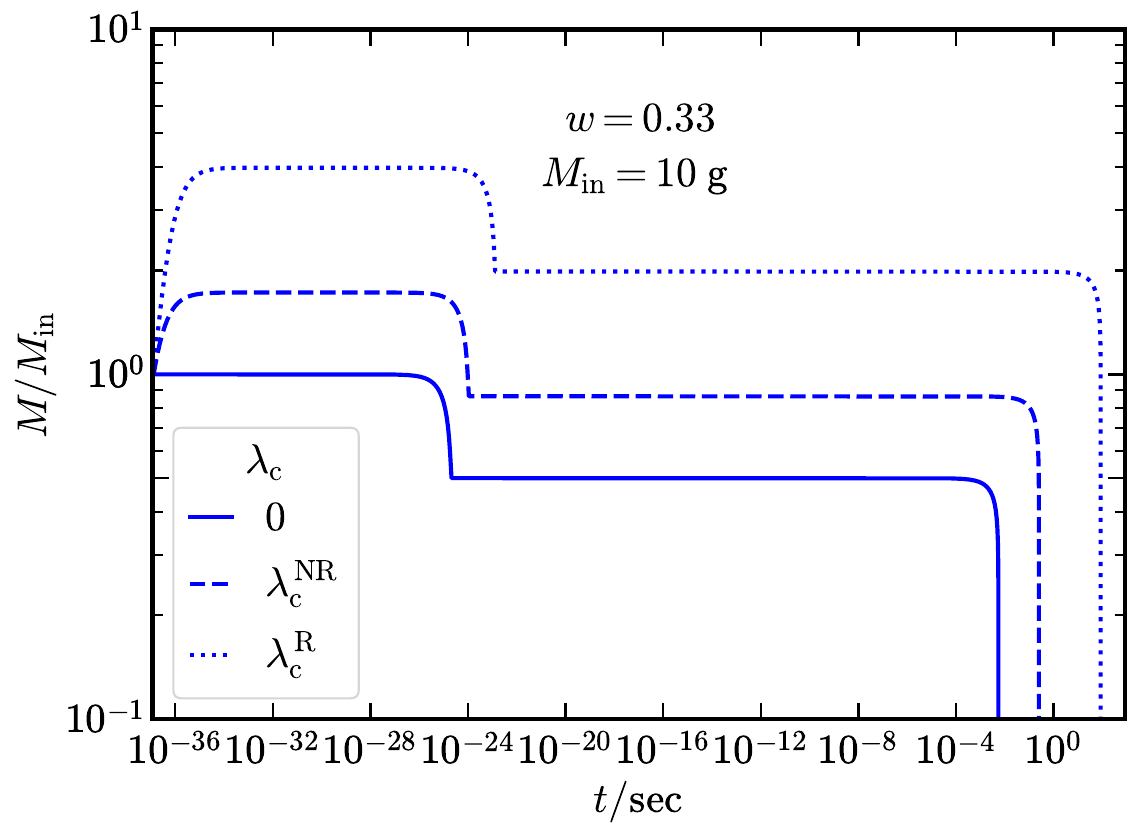}
    \includegraphics[width=0.49\linewidth]{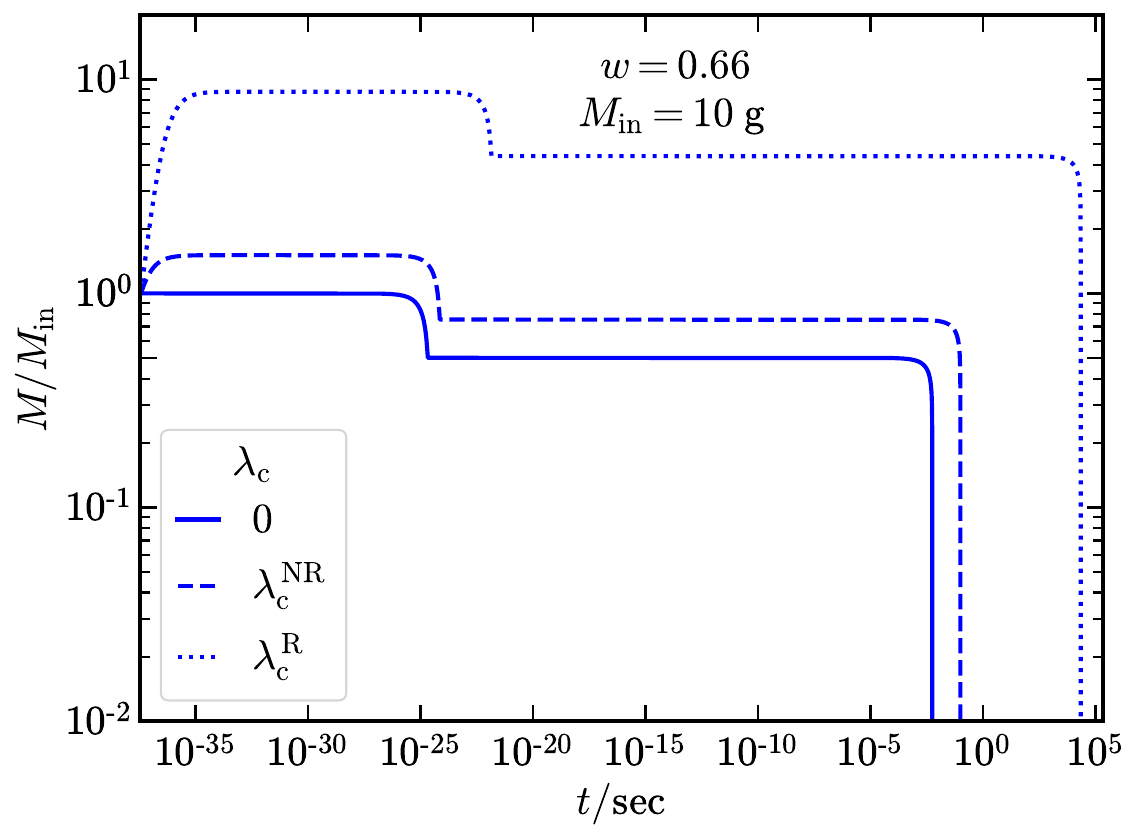}
    \caption{The evolution of the mass of PBHs under the combined effects of accretion and burdened evaporation is shown for two different background equations of state: $w=1/3$ (left panel) and $w=2/3$ (right panel). In each plot, the solid line denotes the evolution of the mass of PBHs without accretion, i.e., purely due to evaporation with memory burden effects. We have fixed the parameter values to $\kappa=2,~q=1/2$. The dashed line represents the evolution when PBHs accrete from a nonrelativistic fluid background, while the dotted line corresponds to accretion from a relativistic fluid.}
    \label{fig:m-acc-evap}
\end{figure*}

\subsection{Accretion of PBHs}

The rate of change of mass of PBHs due to the accretion is given by~\cite{Aguayo-Ortiz:2021jzv,Das:2025vts} 
\begin{align}
    \f{\d M}{\d t}=\f{\lc}{16\pi}\f{M^2}{\Mpl^4}\rho_{\infty},
    \label{eq:massrateacc}
\end{align}
where $\rho_{\infty}$ is the energy density of the background.
The quantity $\lc$ is given by $\lcr=(1/4)w^{-3/2}(1+3w)^{(1+3w)/2w}$ for a relativistic fluid, whereas for a nonrelativistic fluid $\lcnr=(1/4)(\e/w)^{3/2}$~\cite{Ali-Haimoud:2016mbv}.
For a background with EoS $w$, the energy density in terms of the cosmic time is given as $\rho=3\Mpl^2H^2= 4\Mpl^2/[3(1+w)^2t^2]$. Clearly $\lc=0$ indicates that there is no accretion of matter.
Using this behavior of $\rho_\infty$, it is straightforward to integrate Eq.~\eqref{eq:massrateacc} to find the mass of PBHs at any time as~\cite{Das:2025vts}
\begin{align}
    M(t)=\Min \l[1-\f{\lambda_{\rm c}\gamma}{2(1+w)} \f{(3/2)(1+w)\hin(t-\tin)}{(3/2)(1+w)\hin(t-\tin)+1}\r]^{-1}.
    \label{eq:mass-acc1}
\end{align}
A common intuition is that as a PBH survives longer, it continues to grow by accreting surrounding matter or radiation. To understand this quantitatively, let us analyze the behavior described by Eq.~\eqref{eq:mass-acc1}. At early times, accretion is efficient due to high density of the background, but as the Universe expands and $\rho_\infty$ falls, the rate of accretion decreases. Eventually, at a later time ($t\gg\tin$), the accretion saturates when the rate of fractional mass growth becomes equal to the Hubble parameter, i.e., $\dot M/M=H$, and the mass becomes
\begin{align}
    \mac \simeq\Min\l(1-\f{\lambda_{\rm c}\gamma}{2(1+w)}\r)^{-1}.
    \label{eq:mass-acc}
\end{align}
Equation~\eqref{eq:mass-acc} shows that in a radiation-dominated background ($w=1/3$), the total mass gain from relativistic accretion can be estimated as $\mac / \Min \approx 4$. For higher values of the $w$, this ratio increases, and can reach  $\mathcal{O}(100)$ for sufficiently stiff backgrounds. Notably, in the limit of a stiff fluid ($w=1$), the expression formally yields  $\mac / \Min \to \infty$. This apparent divergence arises because accretion requires the black hole to pull matter in faster than the fluid can respond via pressure. In the case of a stiff fluid, pressure disturbances propagate at the speed of light, making the fluid extremely resistant to compression or inward flow. As a result, no steady-state, transonic accretion solution exists, and the standard formalism becomes ill-defined in this limit. Hence, in this work, we focus on two representative background EoS values: $w=1/3$ (radiation domination), and a stiffer case of $w=2/3$, as illustrative toy models\footnote{{ Such a scenario with $w=2/3$ is not just a toy scenario but possible for various models. For instance, consider the inflationary $\alpha$-attractor models~\cite{Odintsov:2016vzz,Ueno:2016dim,Kumar:2015mfa,Eshaghi:2016kne,Dalianis:2018frf,German:2021tqs}. Near the minimum, the potential typically behaves as $V_{\rm min}\sim\phi^{n}$. It can be shown that the corresponding equation of state is given by  $w=(n-2)/(n+2)$~\cite{Garcia:2020wiy}.  For example, $n=2$ and $n=4$ lead to $w=0$ (matter-like) and $w=1/3$ (radiation-like) respectively, while $w=2/3$ corresponds to the case $n=10$.}}.

\color{black}
\subsection{Evaporation of PBHs}

The rate of change of the mass of PBHs due to Hawking evaporation, including the effects of quantum memory burden, is given by~\cite{Hawking:1974rv,Alexandre:2024nuo,Thoss:2024hsr}
\begin{align}
    \f{\d M}{\d t}= -\epsilon\f{\Mpl^4}{M^2}
    S^{-\kappa\Theta(t-\tq)},
\end{align}
where the negative sign reflects the loss of mass due to evaporation. The parameter $\epsilon$ is the graybody factor~\cite{MacGibbon:1990zk,Auffinger:2020afu,Masina:2021zpu,Cheek:2021odj}, which encodes the frequency-dependent transmission probabilities for particles emitted by the PBH evaporation. In the geometric optics limit, it is approximately given by $\epsilon \simeq(27/4)\,g(\Tbh)\pi/480$, where $g(\Tbh)$ is the effective number of relativistic degrees of freedom at the black hole temperature, $\Tbh$, and we have $g(\Tbh)\simeq106.75$.
The factor $S^{-\kappa}$ accounts for the quantum memory burden, which suppresses the evaporation rate once a significant portion of the PBH’s mass has been radiated away.

Initially, PBHs evaporate through the standard semiclassical Hawking radiation. However, once the mass of PBHs reduces to a fraction $q$ of its initial mass $\Min$, memory burden effects become significant, modifying the rate of evaporation and potentially stabilizing the PBH remnant~\cite{Dvali:2012en,Dvali:2020wft,Michel:2023ydf,Wang:2023wsm}. The time at which memory burden starts to dominate is given by 
$\tq=(1-q^3)\tev$, where  $\tev\simeq \Min^3/(3\epsilon\Mpl^4) $ is the time of evaporation of the PBH in a semiclassical process.
The final time of evaporation due to the memory burden is given by $\tevk= [2^\kappa(3+2\kappa)\Mpl]^{-1} \l({q\Min}/{\Mpl}\r)^{3+2\kappa}$~\cite{Haque:2024eyh,Alexandre:2024nuo}.
The mass of the PBHs at any time is now given by
\begin{align}
    M(t)=\begin{cases}
        \Min\l[1-\f{t-\tin}{\tev}\r]^{1/3}& \text{for }t<\tq,\\
        q\Min\l[1-\f{t-\tq}{\tevk}\r]^{1/(3+2\kappa)}& \text{for }t>\tq.
    \end{cases}
\end{align}
We note that the exact value of the memory burden parameter $\kappa$ is currently undetermined, and in this work, we treat it as a free phenomenological parameter. The limiting case $\kappa=0,~q=1$ corresponds to standard Hawking evaporation without any memory burden effects. In principle, the parameter $q$ can take values in the range $0<q<1$. However, following previous discussions in the literature, we adopt the representative value $q=1/2$, for which the memory burden is expected to become relevant~\cite{Alexandre:2024nuo,Thoss:2024hsr}.

\subsection{Total mass evolution of PBHs:}

So far, we have discussed two key physical processes that govern the evolution of the mass of PBHs after their formation: accretion, through which PBHs gain mass by drawing in surrounding matter or radiation, and evaporation, by which they lose mass through the emission of particles. To accurately capture the time evolution of the mass of PBHs, it is essential to incorporate both effects simultaneously.
The total rate of change of the PBH mass is therefore given by the combined equation
\begin{align}
     \f{\d M}{\d t}=\f{\d M}{\d t}\Big\vert_{\rm accretion}
     + \f{\d M}{\d t}\Big\vert_{\rm evaporation}.
\end{align}
At the initial stage of PBH evolution, accretion dominates over evaporation due to the high background energy density. During this phase, PBHs rapidly gain mass until reaching a saturation value $\mac$, beyond which accretion becomes inefficient and evaporation effects begin to take over. The timescale for this accretion-driven growth, denoted by the interval $[\tin,\tac]$, is typically much shorter than the subsequent timescale of evaporation $[\tac,t_{\rm ev}]$. Given this clear separation of timescales, the evolution of the PBH mass can be well approximated analytically by treating $\mac$ as the effective initial mass for the evaporation-dominated phase.

\begin{figure*}
    \centering
    \includegraphics[width=0.49\linewidth]{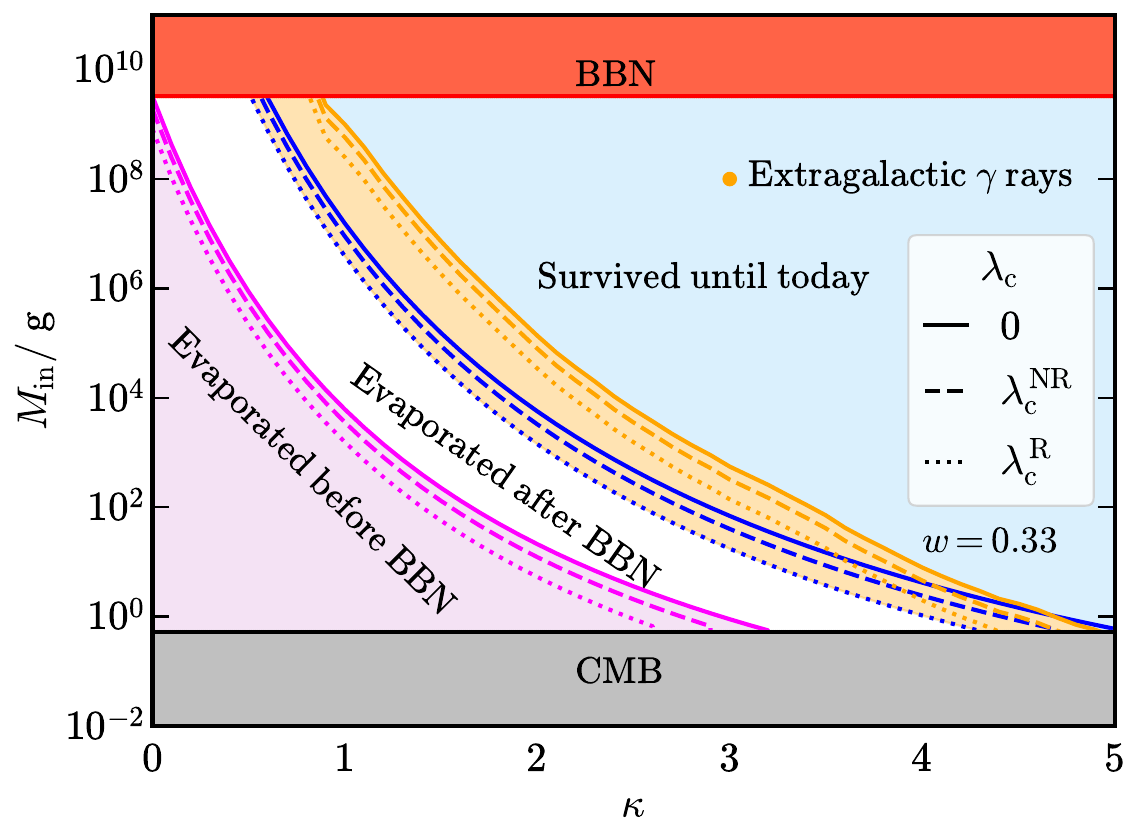}
    \includegraphics[width=0.49\linewidth]{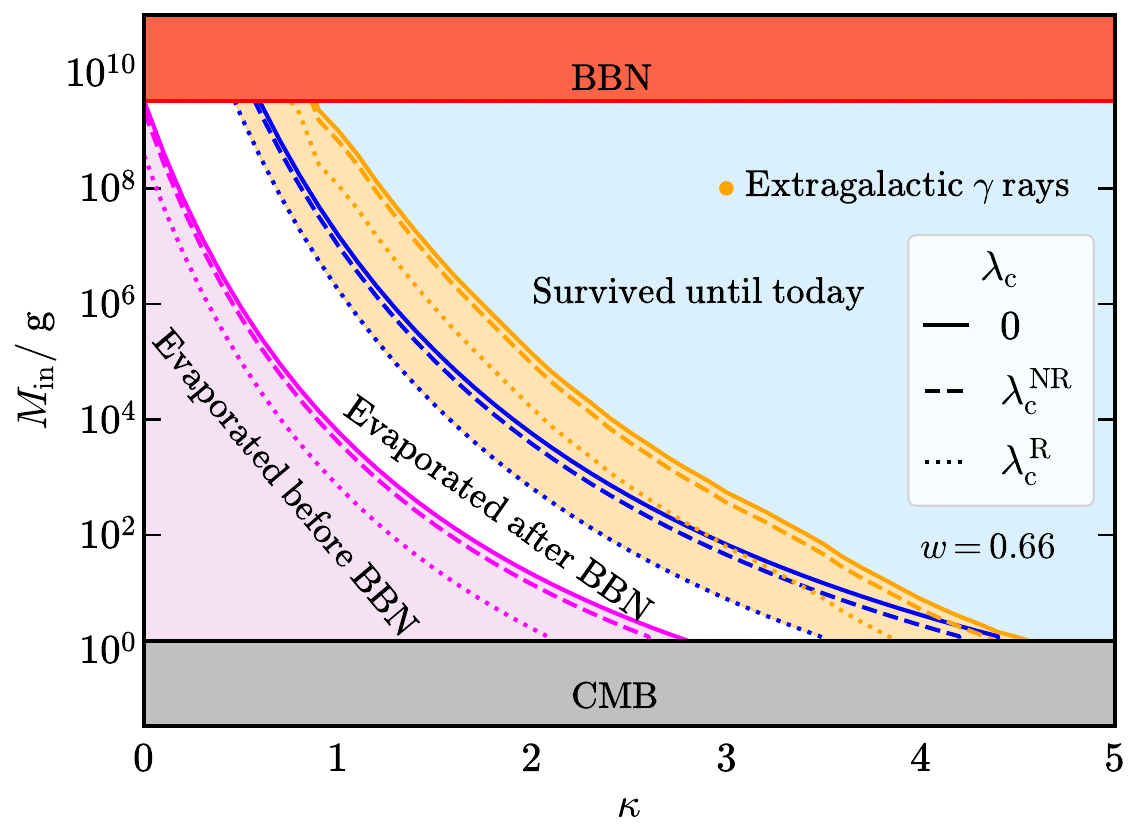}
    \caption{The allowed range of the initial PBH mass $\Min$ as a function of the memory burden parameter $\kappa$ is shown for two background equations of state: $w=1/3$ (left panel) and $w=2/3$ (right panel). The region is bounded from above and below by two key constraints. The gray shaded region at the bottom corresponds to the minimum initial PBH mass allowed by the maximum energy scale of inflation, consistent with the current observational upper bound on the tensor-to-scalar ratio $r=0.036$. 
     The orange shaded region is  constraints from the extragalactic $\gamma$ rays (calculated with observed data of LHAASO~\cite{LHAASO:2023gne} in~\cite{Chianese:2025wrk,Ambrosone:2026djo}).
    The red shaded region indicates values of $\Min$ for which the semiclassical evaporation completes after the onset of BBN. The magenta curves show the upper bound on $\Min$ for which PBHs evaporate completely before BBN, while the blue curves mark the lower bound for which PBHs can survive until the present epoch. The solid, dashed, and dotted curves correspond to scenarios with $\lc=0$, $\lcnr$, and $\lcr$, respectively. 
 The blue region is important in the context of calculating $\fpbh$, and we will also discuss the magenta shaded region in the context of DM produced through the evaporation of PBHs.}
    \label{fig:k-min-1}
\end{figure*}

The complete evolution of the mass of a PBH can now be expressed as
\begin{align}
    M(t)=\begin{cases}
    \Min \Big[1-\f{\lambda_{\rm c}\gamma}{2(1+w)} \times
    \\\quad\f{(3/2)(1+w)\hin(t-\tin)}{(3/2)(1+w)\hin(t-\tin)+1}\Big]^{-1} &\text{for } \tin<t<\tac,
    \\
        \mac\l[1-\f{t-\tac}{\tev}\r]^{1/3}& \text{for } \tac<t<\tq,\\
        q\mac\l[1-\f{t-\tq}{\tevk}\r]^{1/(3+2\kappa)}& \text{for }\tq<t<t_{\rm ev},
    \end{cases}
\end{align}
with the modified expressions of time to be 
\begin{align}
    \tev &\simeq \f{\Min^3}{3\epsilon\Mpl^4}\l(1-\f{\lambda_{\rm c}\gamma}{2(1+w)}\r)^{-3}\\
    \tq &\simeq \f{(1-q^3)\Min^3}{3\epsilon\Mpl^4}\l(1-\f{\lambda_{\rm c}\gamma}{2(1+w)}\r)^{-3}\\
    \tevk &\simeq \f{1}{2^\kappa (3+2\kappa)\epsilon\Mpl} \l(\f{q\Min}{\Mpl}\r)^{(3+2\kappa)}
    \nn\\&\quad
    \l(1-\f{\lambda_{\rm c}\gamma}{2(1+w)}\r)^{-(3+2\kappa)}.\label{eq:tevgen}
\end{align}
Clearly, the expression for the time of evaporation, $\tevk$, given in Eq.~\eqref{eq:tevgen}, represents the most general case. Here, setting $\lc=0$ corresponds to the scenario without accretion, while $\lcnr$ and $\lcr$ denote the cases with nonrelativistic and relativistic accretion, respectively. Furthermore, the choice $\kappa=0$ and $q=1$ recovers the standard case of Hawking evaporation.
In Fig.~\ref{fig:m-acc-evap}, we illustrate the evolution of the mass of PBHs over time. We have chosen the initial mass of the PBH to be $10~\mathrm{g}$, and the two panels correspond to different background EoS, namely $w=1/3$ and $w=2/3$, respectively. The solid, dashed, and dotted lines represent the evolution of the PBH mass in the absence of accretion ($\lc=0$), with nonrelativistic accretion ($\lcnr$), and with relativistic accretion ($\lcr$), respectively. We have fixed the parameters characterizing the memory burden to $\kappa=2$ and $q=1/2$. It can be seen that the initial mass gain, $\mac/\Min$, due to relativistic accretion can reach $\sim 4$ for a background with $w=1/3$, and can be as large as $\sim 10$ for $w=2/3$. The increase in mass due to accretion consequently extends the lifetime of the PBHs, with relativistic accretion leading to an enhancement of the lifetime by several orders of magnitude.

In this work, we focus on two scenarios: one in which the PBH completely evaporates before the onset of BBN, and another in which the PBH survives until the present epoch.
By taking into account both accretion and evaporation, the former case imposes a constraint on the initial mass of the PBH, for which the PBH would evaporate entirely before BBN to be 
\begin{align}
    \Min^{\rm BBN} &\leq \f{\Mpl}{q}\l(1-\f{\lambda_{\rm c}\gamma}{2(1+w)}\r)\nn\\&\quad (2^\kappa(3+2\kappa)\epsilon\tbbn\Mpl)^{1/(3+2\kappa)}.
    \label{eq:mmax-bbn}
\end{align}
From the relation in Eq.~\eqref{eq:mmax-bbn}, we see that for evaporation purely via Hawking radiation, the maximum initial mass of PBHs that would evaporate before BBN is approximately $\sim 1.6\times 10^9~\mathrm{g}$ in the absence of accretion. This value decreases to $\sim 9.4\times 10^8~\mathrm{g}$ and $\sim 4\times 10^8~\mathrm{g}$ for nonrelativistic and relativistic accretion, respectively, in a radiation-dominated background.
For the scenario with a memory burden characterized by parameters $\kappa=2$ and $q=1/2$, the maximum allowed initial mass reduces significantly to $\sim 21~\mathrm{g}$, $12~\mathrm{g}$, and $5~\mathrm{g}$ for $\lc=0$, $\lcnr$, and $\lcr$, respectively.
Moreover, a different background EoS affects these constraints in the presence of accretion; for instance, with $w=2/3$, the maximum allowed mass for relativistic accretion further decreases to $\sim 2.5~\mathrm{g}$.
In a similar fashion, the 
minimum mass of a PBH to survive the present epoch is given by 
\begin{align}
    \Min^0\geq \f{\Mpl}{q}\l(1-\f{\lambda_{\rm c}\gamma}{2(1+w)}\r)(2^\kappa(3+2\kappa)\epsilon t_0\Mpl)^{1/(3+2\kappa)}.
\end{align}
We know that for standard Hawking evaporation without accretion, the minimum mass of PBHs that can survive until today is approximately $10^{15}~\mathrm{g}$.
When including the effects of a memory burden with parameters $\kappa=2$ and $q=1/2$, this minimum mass shifts significantly, becoming $\sim 5600~\mathrm{g}$, $3200~\mathrm{g}$, and $1400~\mathrm{g}$ for the cases $\lc=0$, $\lcnr$, and $\lcr$, respectively.
Moreover, for a background EoS with $w=2/3$ and relativistic accretion, the minimum mass required for PBHs to survive until the present epoch further decreases to $\sim 640~\mathrm{g}$.
In Fig.~\ref{fig:k-min-1}, we illustrate the various constraints on the initial mass of PBHs as a function of the memory burden parameter $\kappa$. We consider two values of the background EoS: $w=1/3$, shown in the left panel, and $w=2/3$, shown in the right panel. In each plot, the gray shaded regions correspond to the minimum allowed initial mass of PBHs, set by the highest permissible energy scale of inflation inferred from the tensor-to-scalar ratio $r=0.036$~\cite{BICEP:2021xfz}. The red shaded regions indicate the upper limit on the initial mass above which Hawking evaporation would occur only after BBN, while the magenta lines mark the maximum initial mass for which PBHs would completely evaporate before BBN. In contrast, the blue lines denote the minimum initial mass required for PBHs to survive until the present epoch. The solid, dashed, and dotted lines correspond to the cases with $\lc=0$, $\lcnr$, and $\lcr$, respectively. We observe that the constraints arising from relativistic accretion differ significantly from the other cases, with the disparity becoming more pronounced for a stiffer background EoS.
  The orange shaded region is the constraints where $\fpbh<1$, coming from the extragalactic $\gamma$ rays. Note that the constraints are calculated with observed data of LHAASO~\cite{LHAASO:2023gne} with the methodology given in~\cite{Chianese:2025wrk,Ambrosone:2026djo}.
 Discussion about various constraints on $\fpbh$ is given in Sec.~\ref{sec:dm-stable}.

Let us now discuss the possibility that the energy density of PBHs dominates over the energy density of the background before the PBHs evaporate.
As the Universe expands, the energy densities evolve with the scale factor $a$ as
\begin{align}
    \rhobh=\rhobh^{\rm in}\l(\f{\ain}{a}\r)^{3},\quad
    \rhow=\rhow^{\rm in}\l(\f{\ain}{a}\r)^{3(1+w)},
\end{align}
where the sub/superscript `in' denotes the quantities at the time of the formation of PBHs.
For $w > 0$, the energy density of PBHs redshifts more slowly than the energy density of the background, implying that even a subdominant initial PBH population could eventually come to dominate the total energy density of the Universe before evaporation.
We further know that the scale factor behaves with the cosmic time as $a\propto t^{2/[3(1+w)]}$. Using this, we get the time when the energy density of PBHs starts dominating the background energy density as 
\begin{align}
    \tbh=\tin (q\beta)^{-(1+w)/(2w)}.
\end{align}
Moreover, for the PBH domination to take place, the time of the domination needs to be before the time of the evaporation of PBHs, $\tbh\leq \tevk$.
These conditions lead to the critical value of the initial abundance of PBHs, $\beta$, above which there will be PBH domination.  The critical value of $\beta$ is given by 
\begin{align}
    \betacr=\f{1}{q}\l(\f{\tin}{\tevk}\r)^{\f{2w}{1+w}}&=
    \f{1}{q} \l[\f{2^\kappa(3+2\kappa)\epsilon}{6\pi\gamma(1+w)q} \l(\f{\Mpl}{q\Min}\r)^{2(1+\kappa)}\r]^{\f{2w}{1+w}}
    \nn\\&\quad 
    \l(1-\f{\lc\gamma}{2(1+w)}\r)^{\f{2w(3+2\kappa)}{1+w}}.
\end{align}
Again, we get for $\kappa=2$, $q=1/2$ and $w=1/3$,
\begin{align*}
    \betacr^{\lc=0}&=\f{9.67\times10^{-15}}{\Min^3},~
    \betacr^{\lcnr}=\f{1.5\times10^{-15}}{\Min^3},~
    \nn\\&\quad
    \betacr^{\lcr}=\f{7.5\times10^{-17}}{\Min^3},~
\end{align*}
whereas for $w=2/3$
\begin{align*}
    \betacr^{\lc=0}&=\f{9.1\times10^{-24}}{\Min^{4.8}},~
    \betacr^{\lcnr}=\f{9.1\times10^{-25}}{\Min^{4.8}},~
    \nn\\&\quad
    \betacr^{\lcr}=\f{4.2\times10^{-29}}{\Min^{4.8}}.~
\end{align*}
In the semiclassical evaporation $(q=1,~\kappa=0)$, $\betacr$ is given by for $w=1/3$
\begin{align*}
    \betacr^{\lc=0}&=\f{7.42\times10^{-6}}{\Min},~
    \betacr^{\lcnr}=\f{3.27\times10^{-6}}{\Min},~
    \nn\\&\quad
    \betacr^{\lcr}=\f{9.28\times10^{-7}}{\Min}.~
\end{align*}
and for $w=2/3$
\begin{align*}
    \betacr^{\lc=0}&=\f{2.26\times10^{-9}}{\Min^{1.6}},~
    \betacr^{\lcnr}=\f{8.46\times10^{-10}}{\Min^{1.6}},~
    \nn\\&\quad
    \betacr^{\lcr}=\f{1.17\times10^{-11}}{\Min^{1.6}}.~
\end{align*}
In deriving the expression for $\betacr$, we have assumed the possibility of PBH domination, wherein the Universe undergoes a sequence of phases: $w$-domination $\rightarrow$ PBH domination $\rightarrow$ radiation domination. A more intricate scenario can arise if the inflaton is strongly coupled to SM particles and decays before PBH domination, leading to an intermediate phase of radiation domination prior to PBH domination ($w$-domination $\rightarrow$ radiation domination $\rightarrow$ PBH domination $\rightarrow$ radiation domination). Such a scenario is often called as two-phase reheating, where the Universe is dominated by radiation twice. There are some subtleties with this point, which we clarify below:
\begin{itemize}
    \item 
 A stronger coupling of the inflaton to SM particles results in a faster decay of the inflaton. After the formation of PBHs during a $w$-dominated phase, the inflaton may decay into radiation while the PBHs still survive. Hence, the decay of the inflaton leads to a phase of radiation domination. Since the energy density of radiation redshifts faster than that of PBHs, the PBHs can eventually dominate the energy density of the Universe before they evaporate. In such a scenario, a second phase of reheating can take place due to the radiation produced during the evaporation of PBHs.
 This scenario is understood as 
 $w$-domination $\rightarrow$ radiation domination $\rightarrow$ PBH domination $\rightarrow$ radiation domination.
\item
We clarify that during the $w$-dominated phase, if $w = 1/3$, the expansion behaves as radiation domination, even if the radiation component is not associated with SM particles.
In this case, the original four-phase scenario effectively reduces to a three-phase sequence:
radiation domination $ \rightarrow$ PBH domination $ \rightarrow$ radiation domination.
Notably, this setup can also be interpreted as a form of double radiation domination, in which the strength of the inflaton–SM particle coupling does not play a significant role.
\end{itemize}
Hence, to summarize, let us note that the expression of $\betacr$ is derived in this work, assuming a sufficiently small coupling of the inflaton to radiation where inflaton is decaying after PBH domination. This can be used for a strong coupling case only if $w=1/3$ during the inflaton-dominated phase.

\section{Production of dark matter from evaporation of PBHs \label{sec:dm-evap}}

In this section, we discuss the production of DM from the evaporation of PBHs. We specifically restrict our analysis to the mass range of PBHs that ensures they would have evaporated completely before the onset of BBN. We begin by calculating the number of DM particles produced from the evaporation of a single PBH.
It is important to note that both the mass and spin of a black hole significantly affect the production rate of particles generated during its evaporation. The emission spectra and graybody factors depend sensitively on the spin of the emitted species as well as the angular momentum of the black hole itself. However, for simplicity, in this work we restrict our analysis to the case of non-rotating, spin-zero Schwarzschild black holes.
Under this assumption, the emission rate of a particle species $j$ with mass $m_j$ and internal degrees of freedom $g_j$, per unit time and per unit energy interval, is given by~\cite{RiajulHaque:2023cqe,Haque:2024eyh}
\begin{align}
\f{\d^2 N_{j}}{\d E\d t} &
= \f{27}{4} \f{g_j}{32 \pi^3}\f{(E/\Tbh)^2}{\exp(E/\Tbh)\pm 1}S^{-\kappa\,\Theta(t-\tq)}
\label{eq:dndtde}
\end{align}
where the sign $\pm$ corresponds to fermions and bosons, respectively. The factor $S^{-\kappa}$ ensures that the flux of the emitted particles is also modified by the memory burden, reflecting the change in the evaporation rate induced by the parameter $\kappa$.
The number of particles emitted per unit time due to the evaporation can be calculated by integrating Eq.~\eqref{eq:dndtde}, over the energy, and it is given by 
\begin{align} 
\f{\d N_{1j}}{\d t} &=\f{27}{4} \f{\xi g_j  \zeta(3)}{16\pi^3} \f{\Mpl^2}{M}, \label{eq:dn1nt}\\
\f{\d N_{2j}}{\d t} &=\f{27}{4} \f{\xi g_j  \zeta(3)2^k}{16\pi^3} \f{\Mpl^{2+2\kappa}}{M^{1+2\kappa}}
\label{eq:dn2dt}
\end{align}
where $\xi=1$ for bosons and $\xi=3/4$ for fermions.
The quantities $N_{1j}$ and $N_{2j}$ denote the number of particles of species $j$ produced during the semiclassical first phase and the subsequent quantum memory burden phase of evaporation, respectively.

To calculate the total number of particles produced through evaporation, one integrates the emission rate $\d N_j/\d t$ over the lifetime of the PBH. In this context, two distinct scenarios arise:
if the mass of the emitted particle satisfies $m_j < \Tbh^{0}$, where $\Tbh^{0}$ is the Hawking temperature of the PBH when the PBH starts evaporating, then emission occurs throughout the entire lifetime of PBHs. On the other hand, if $m_j > \Tbh^{0}$, production only begins at a later time $t_j$ when the PBH temperature rises to satisfy $\Tbh(t_j) = m_j$.
One can find the time $t_j$ to be 
\begin{align} 
t_j=
\begin{cases}
{\tev}\Big[1-\f{\Mpl^6}{\Min^3 m_j^3} \l(1-\f{\lambda_{\rm c}\gamma}{2(1+w)}\r)^3\Big]& \\
\hspace{5.5cm}\text{for}~t_j<\tq,
\\
\\
\tevk\l[1-\l(\f{\Mpl^2}{q\Min m_j}\r)^{3+2k} \l(1-\f{\lambda_{\rm c}\gamma}{2(1+w)}\r)^{3+2\kappa}\r]& \\
\hspace{5.5cm}\text{for}~t_j>\tq\,,
\end{cases}
\label{eq:ti}
\end{align}  
where $\tq$ is the time after which memory burden starts dominating, defined before.

For the case of $m_j<\tbh^{\rm acc}$, the DM will be produced throughout the time $[\tac,\tevk]$. 
One can calculate the number of particles produced in the semiclassical and memory-burdened phase, $N_{ij}$ and $N_{2j}$, by integrating Eq.~\eqref{eq:dn1nt} in the time interval $[\tac,\tq]$ and Eq.~\eqref{eq:dn2dt} over the time interval $[\tq,\tevk]$ respectively. For the detailed derivation regarding this, see our previous work~\cite{Haque:2024eyh}. 
In this paper, we also showed that for complete evaporation of PBHs, the total number of emitted particles is constant and does not depend on the rate of emission. 
The total number of DM particles with $m_j<\tbh^{\rm acc}$, emitted from a single BH is thus given by 
\begin{align} 
N_j=N_{1j}+N_{2j}=\f{15 \xi g_j\zeta(3)}{g_{\ast}(\Tbh)\pi^4}\f{\Min^2}{\Mpl^2}\l(1-\f{\lambda_{\rm c}\gamma}{2(1+w)}\r)^{-2}\,,
\label{Eq:mjlTbh}
\end{align}
which is the same as the total number of particles emitted from the Hawking evaporation, as expected. 
Similarly, for the case $m_j>\Tbh^{\rm acc}$,
two scenarios can arise: $t_j<\tq$ and $t_j>\tq$. 
For $t_j<\tq$, $N_{ij}$ and $N_{2j}$ can be calculated by integrating Eqs.~\eqref{eq:dn1nt} and \eqref{eq:dn2dt} over the time intervals $[t_j,\tq]$ and $[\tq,\tevk]$ respectively. 
Whereas for $t_j>\tq$, $N_{ij}=0$ and $N_{2j}$ can be calculated by integrating over $[t_j,\tevk]$.
In both cases, after complete evaporation, we get the total number of DM particles emitted to be
\begin{align} 
N_j=N_{1j}+N_{2j}=\f{15\xi g_j \zeta(3)}{g_{\ast}(\Tbh)\pi^4}\f{\Mpl^2}{m_j^2}.
\label{Eq:mjgTbh}
\end{align} 
The relic density of DM today is given by  the ratio between the energy density of DM to the critical energy density of the Universe as~\cite{Mambrini:2021cwd}
\begin{align}
    \Omega_j= \f{\rho_j^0}{\rho_{\rm c}^0}
    =\f{S_0}{\rho_{\rm c}^0}\f{n_j^{\rm re}}{S_{\rm re}}m_j
\end{align}
where $S_0 = 2.23 \times 10^{-38}~\text{GeV}^3$ is the entropy density of the Universe today, and $\rho_{\rm c}^0 = 5.15 \times 10^{-47}~\text{GeV}^4$ is the current critical energy density. The quantity $n_j^{\rm re}$ denotes the number density of the emitted DM particles at the end of reheating, and since the Universe evolves adiabatically after reheating, the ratio $n_j/S$ remains conserved.
For a radiation-dominated Universe, the entropy density is related to the temperature as
$S_{\rm re}=(25\pi^2/45)g_s^{\rm re} \Tre^3$, where $g_s^{\rm re}$ is the effective number of relativistic degrees of freedom contributing to entropy at the end of reheating.
As mentioned before, we shall consider two scenarios: $\beta>\betacr$, when the energy density of PBHs dominates the energy density of the background, and $\beta<\betacr$ for which there will not be any PBH domination.

\subsection{The case with $\beta>\betacr$}

\begin{figure*}
    \centering
    \includegraphics[width=0.49\linewidth]{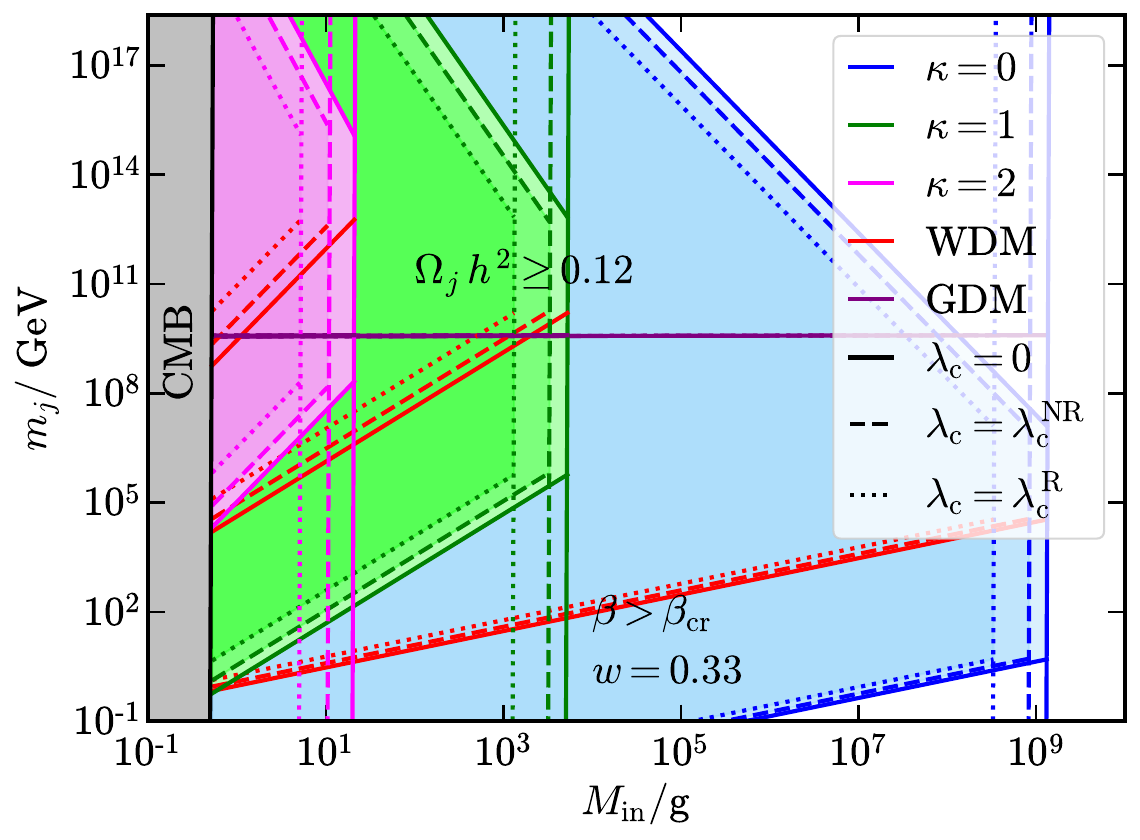}
    \includegraphics[width=0.49\linewidth]{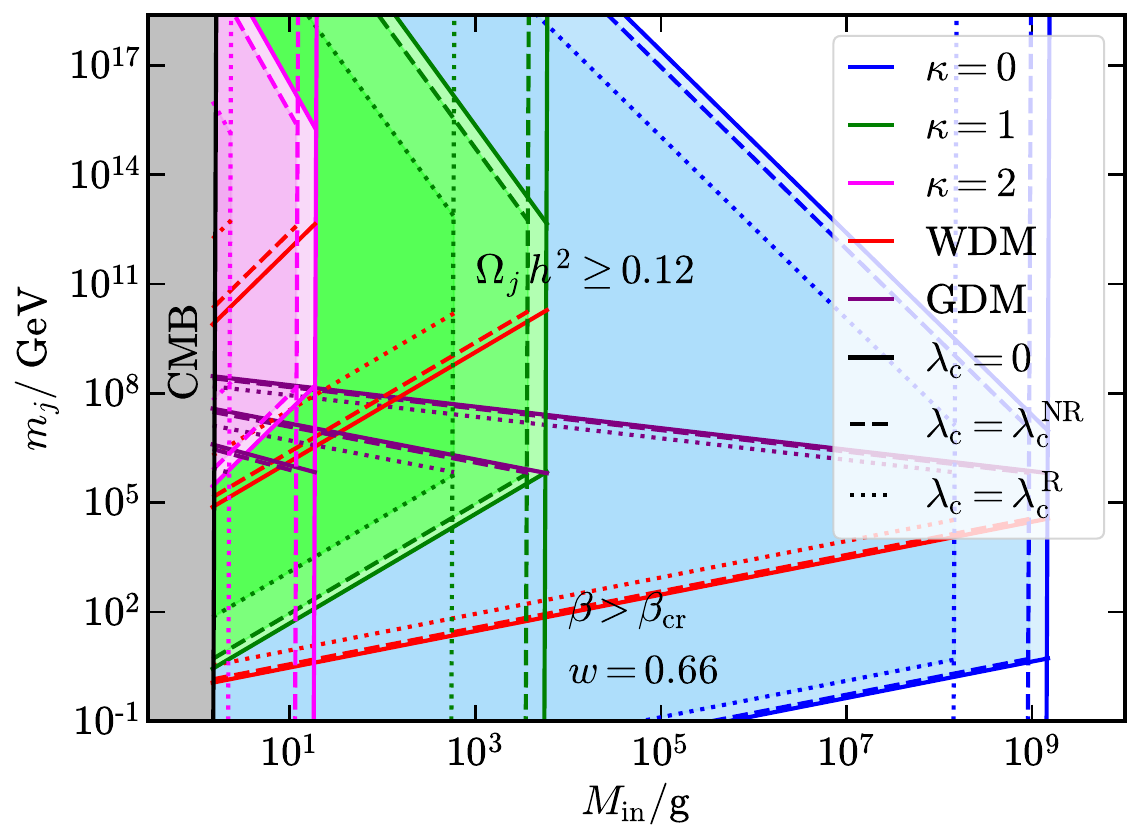}
    \caption{The mass of DM produced from the evaporation of PBHs is shown as a function of the initial mass of PBHs for the case $\beta > \betacr$ { to produce a particular $\Omega_j\,h^2$ (using Eqs.~\eqref{eq:ojs1} and~\eqref{eq:ojl1}). The lines represent the combinations of $\Min-m_j$ for which $\Omega_j\,h^2=0.12$}. We consider two background equations of state: $w=1/3$ displayed in the left panel, and $w=2/3$ in the right panel.
The gray shaded region corresponds to the minimum initial mass of PBHs set by the maximum energy scale of inflation, consistent with the current observational upper bound on the tensor-to-scalar ratio $r=0.036$. The shaded colored regions indicate the parameter space that leads to an overproduction of DM, i.e. $\Omega_j,h^2 > 0.12$. The vertical lines mark the maximum initial mass of PBHs that ensures complete evaporation of PBHs before BBN.
For both these plots, we fix $q=1/2$. The blue, green, and magenta lines correspond to memory burden parameters $\kappa=0$, $1$, and $2$, respectively, while the solid, dashed, and dotted lines represent the cases $\lc=0$, $\lcnr$, and $\lcr$, respectively.
The red lines represent the lower bounds on the DM mass, derived from warm DM constraints.  And the purple lines correspond to the constraints coming from gravitationally produced DM.} 
    \label{fig:dm-bg}
\end{figure*}

In this case, the energy density of the PBHs dominates the background. The reheating is governed solely by the evaporation of the PBHs. Hence, at the time of evaporation, the Hubble parameter behaves as $H_{\rm ev} = 2/(3\tevk)$. Further, as the reheating occurs due to the evaporation of PBHs, we get $\rhor^{\rm re} = \rhobh^{\rm re} = \rhobh^{\rm ev}$.
The number density of DM is related to the number density of PBHs as $n_j^{\rm ev} = N_j \times \nbh^{\rm ev}$, where $N_j$ is the number of particles emitted from the evaporation of a single PBH, and $\nbh^{\rm ev}$ is the number density of PBHs at the time of evaporation. The quantity $\nbh^{\rm ev}$ can be obtained as
    \begin{align}
        \nbh^{\rm ev}=\f{4}{3}\Mpl^32^{2\kappa}(3+2\kappa)^2\epsilon^2
        \l(\f{\Mpl}{q\mac}\r)^{7+4\kappa}.
        \label{eq:nbhevg}
    \end{align}
Finally using the expression of $\nbh^{\rm ev}$ from Eq.~\eqref{eq:nbhevg} and the expressions of $N_j$ from Eqs.~\eqref{Eq:mjlTbh} and \eqref{Eq:mjgTbh}, we obtain the relic abundance of DM for 
 $m_j<\Tbh^{\rm acc}$ to be
\begin{align}
    \Omega_j\,h^2=2.43\times 10^5\f{\xi g_j}{q^2}
    \l[2^\kappa(3+2\kappa)\r]^{1/2}
    \l(\f{\Mpl}{q\mac}\r)^{\f{1+2\kappa}{2}} \f{m_j}{\text{GeV}},
    \label{eq:ojs1}
\end{align}
and for $m_j>\Tbh^{\rm acc}$,
\begin{align}
    \Omega_j\,h^2=1.4\times 10^{42}{\xi g_j}
    \l[2^\kappa(3+2\kappa)\r]^{1/2}
    \l(\f{\Mpl}{q\mac}\r)^{\f{5+2\kappa}{2}} \f{\text{GeV}}{m_j}.
    \label{eq:ojl1}
\end{align}
In Fig.~\ref{fig:dm-bg}, we show the mass of DM produced from the evaporation of PBHs as a function of the initial mass of PBHs, for $w=1/3$ (left panel) and $w=2/3$ (right panel). The blue, green, and magenta lines correspond to memory burden parameters $\kappa=0$, $1$, and $2$, respectively. The shaded regions indicate an overproduction of DM, where $\Omega_j,h^2 > 0.12$. The vertical lines mark the maximum initial masses of PBHs that can fully evaporate before the onset of BBN.
We observe that as the memory burden parameter $\kappa$ increases, the allowed parameter space in the $m_j-\Min$ plane becomes progressively smaller. Additionally, the solid, dashed, and dotted lines represent the cases of $\lc=0$, $\lcnr$, and $\lcr$, respectively. From both panels, it is evident that the parameter space changes significantly in the case of relativistic accretion.

\subsection{The case with $\beta<\betacr$}

In this case, the energy density of PBHs never dominates over the background energy density of the Universe. Several scenarios can arise in this context. For instance, PBHs may form and evaporate entirely during a phase dominated by the same background EoS. Alternatively, PBHs could form in a background with a general EoS and subsequently evaporate after the inflaton decays into radiation, leading to the epoch of radiation domination.
As already stated, in this work we focus on the scenario where the coupling between the inflaton and radiation is small, resulting in an extended period of reheating. We also assume that reheating persists up to the onset of BBN. Additionally, we concentrate on PBHs that evaporate before BBN, thereby focusing on the first scenario described above.
Let us begin with the case where the background is dominated by radiation. In this situation, the number density of PBHs at the time of evaporation is given by
\begin{align}
    \nbh^{\rm ev}&=3\beta\l(\f{\pi\gamma}{2}\r)^{1/2}
    [2^\kappa(3+2\kappa)\epsilon]^{3/2} \Mpl^3 \nn\\&\quad
    \l(\f{\Mpl}{\Min}\r)^{3/2}
    \l(\f{\Mpl}{q\mac}\r)^{\f{3}{2}(3+2\kappa)}.
    \label{eq:nbhev2}
\end{align}
Further, during radiation domination using the relation between the energy density and temperature, $\rhobh^{\rm ev}=q\mac\nbh^{\rm ev}=\alpha\Tev^4$, we obtain that the temperature at the time of evaporation is given by
\begin{align}
    \Tev= \l(\f{3}{4\alpha}\r)^{1/4}[2^\kappa(3+2\kappa)\epsilon]^{1/2}\Mpl 
    \l(\f{\Mpl}{q\mac}\r)^{\f{3+2\kappa}{2}}.
    \label{eq:Tev2}
\end{align}
Using the relations given in Eqs.~\eqref{eq:nbhev2} and \eqref{eq:Tev2}, we obtain the relic density of DM 
for $m_j<\Tbh^{\rm acc}$ to be
\begin{align}
    \Omega_j\,h^2=2.1\times 10^5{\xi g_j}\beta
    \l(\f{\Min}{\Mpl}\r)^{1/2}
    \l(1-\f{3\lambda_{\rm c}\gamma}{8}\r)^{-2}
     \f{m_j}{\text{GeV}}.
\end{align}
Also for $m_j>\Tbh^{\rm acc}$ we have
\begin{align}
    \Omega_j\,h^2=1.28\times 10^{42}{\xi g_j}\beta
    \l(\f{\Mpl}{\Min}\r)^{3/2}
    \f{\text{GeV}}{m_j}.
\end{align}
\begin{figure*}
    \centering
    \includegraphics[width=0.49\linewidth]{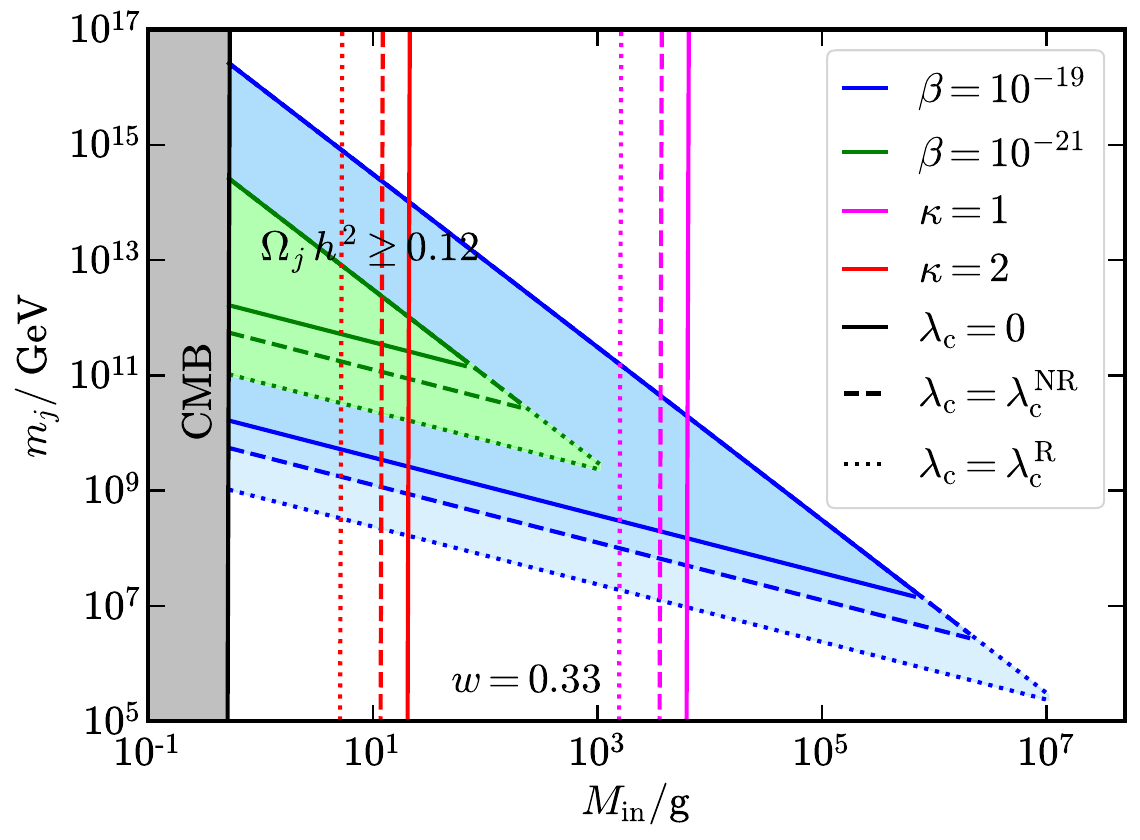}
    \includegraphics[width=0.49\linewidth]{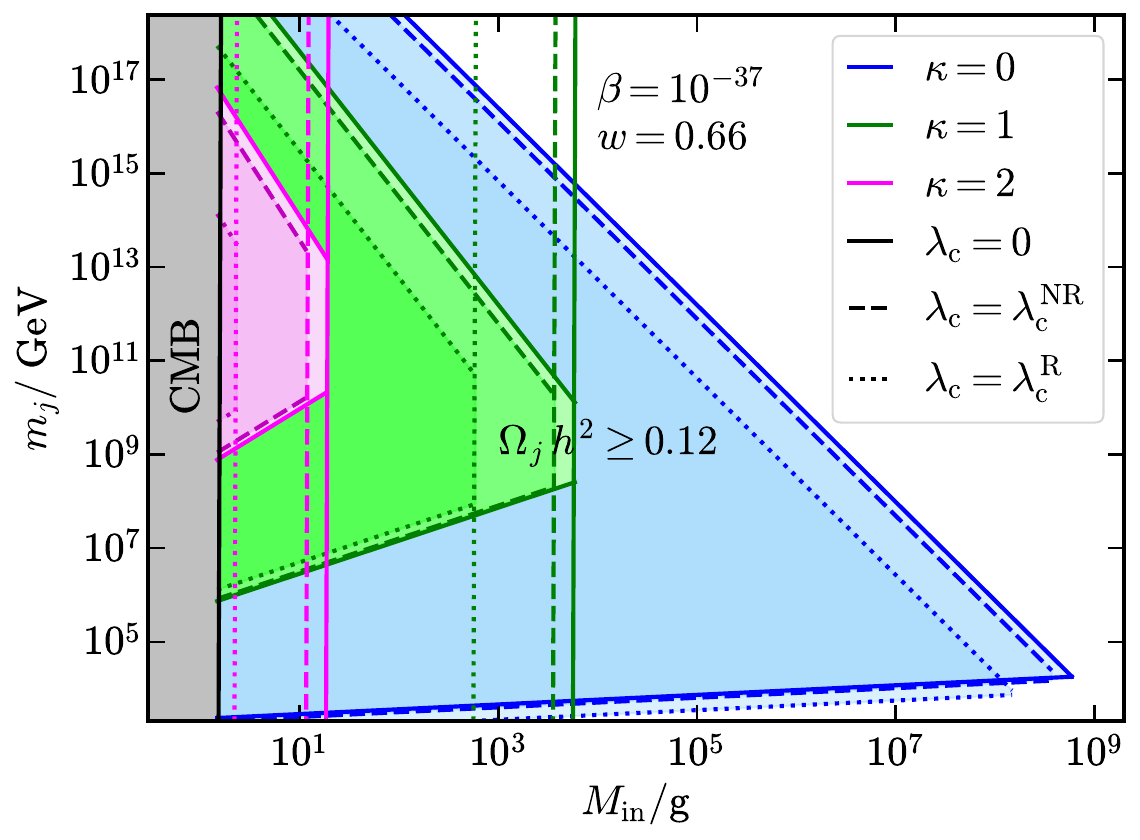}
    \caption{The mass of DM produced from the evaporation of PBHs is shown as a function of the initial mass of PBHs for the case $\beta <\betacr$. We consider two background equations of state: $w=1/3$, displayed in the left panel, and $w=2/3$, shown in the right panel.
As before, the gray shaded region corresponds to the minimum initial mass of PBHs set by the maximum energy scale of inflation. The shaded colored regions indicate the parameter space that leads to an overproduction of DM, i.e. $\Omega_j,h^2 > 0.12$.
In the left panel for $w=1/3$, we present results for two values of $\beta$, where the blue and green lines correspond to $\beta=10^{-19}$ and $\beta=10^{-21}$, respectively. The vertical magenta and red lines denote the maximum initial masses allowed for complete evaporation of PBHs before BBN, for memory burden parameters $\kappa=1$ and $2$, respectively. Throughout, we fix $q=1/2$.
In the right panel, for $w=2/3$, we set $\beta=10^{-37}$. Here, the blue, green, and magenta lines correspond to $\kappa=0$, $1$, and $2$, respectively. Finally, in both panels, the solid, dashed, and dotted lines represent the cases with $\lc=0$, $\lcnr$, and $\lcr$, respectively.
}
    \label{fig:dm-bl}
\end{figure*}
In Fig.~\ref{fig:dm-bl}, we illustrate on the left panel the parameter space of the mass of the emitted DM versus the initial mass of PBHs. We have chosen two different values of $\beta$, namely $10^{-19}$ (shown in blue) and $10^{-21}$ (shown in green). The shaded regions indicate where there is an overproduction of DM, i.e. $\Omega_j\,h^2 > 0.12$. We observe that even a slight change in the parameter $\beta$ significantly affects the allowed parameter space.
From the relation for $\Omega_j,h^2$ discussed earlier, it is clear that the memory burden parameter $\kappa$ does not substantially modify the parameter space, except by restricting the maximum initial mass of PBHs that can evaporate before BBN. This is represented in the plot by the vertical magenta and red lines corresponding to $\kappa=1$ and $2$, respectively. The solid, dashed, and dotted lines denote the cases with $\lc=0$, $\lcnr$, and $\lcr$, respectively. We also see that accretion significantly impacts the parameter space in the $m_j-\Min$ plane, reducing the allowed regions, particularly in the case of relativistic accretion.

Next, we consider the case where the background is governed by an EoS parameter $w$ different from that of radiation. Given our assumption of a weak coupling between the inflaton and radiation, we focus on the scenario where reheating is driven by the evaporation of PBHs. In this context, PBHs may evaporate during the period of $w$-domination, thereby injecting radiation into the Universe. For a stiffer background with $w > 1/3$, the energy density of the background fluid redshifts faster than the energy density of the radiation produced by the evaporation of PBHs. As a result, after a certain time, the energy density of radiation surpasses that of the background, effectively reheating the Universe.   
Hence, for a background with $w$ domination, the Hubble parameter at the time of evaporation is given by $H_{\rm ev}=2/[3(1+w)\tevk]$. Further, the energy density of PBHs at the time of evaporation is given by 
\begin{align}
    \rhobh^{\rm ev}=48\pi^2\beta\Mpl^4
    \l[\f{2^\kappa(3+2\kappa)\epsilon\gamma^w}{6\pi(1+w)}\r]^{\f{2}{1+w}} \l(\f{\Mpl}{\Min}\r)^{\f{1+3w}{1+w}}.
\end{align}
At the point of evaporation, we have $\rhobh^{\rm ev}=\rhor^{\rm ev}=\alpha\Tev^4$. Using this, we obtain
\begin{align}
    \f{\nbh^{\rm ev}}{\Tev^3}=\f{(\alpha^3\rhobh^{\rm ev})^{1/4}}{q\mac}.
\end{align}
Finally, we get an abundance of DM 
for $m_j<\Tbh^{\rm acc}$ 
\begin{align}
    \Omega_j\,h^2&=4.8\times 10^5\f{\xi g_j}{q^2}
    \l[\f{2^\kappa(3+2\kappa)\epsilon\gamma^w}{6\pi(1+w)}\r]^{\f{1}{2(1+w)}}\beta^{1/4}
    \nn\\&\quad
    \l(\f{\Mpl}{\Min}\r)^{\f{1+3w}{4(1+w)}}
     \l(\f{\Mpl}{q\mac}\r)^{\f{1-5w+4\kappa}{4(1+w)}}
     \f{m_j}{\text{GeV}},
\end{align}
and for $m_j>\Tbh^{\rm acc}$ 
\begin{align}
    \Omega_j\,h^2&=2.9\times 10^{42}{\xi g_j}
    \l[\f{2^\kappa(3+2\kappa)\epsilon\gamma^w}{6\pi(1+w)}\r]^{\f{1}{2(1+w)}}\beta^{1/4}
    \nn\\&\quad
    \l(\f{\Mpl}{\Min}\r)^{\f{1+3w}{4(1+w)}}
     \l(\f{\Mpl}{q\mac}\r)^{\f{9+3w+4\kappa}{4(1+w)}}
    \f{\text{GeV}}{m_j}.
\end{align}
In Fig.~\ref{fig:dm-bl}, on the right panel, we illustrate the parameter space between the mass of emitted DM and the initial mass of PBHs, for the case where PBHs form and evaporate in a background with EoS $w = 2/3$. We assume a minimal reheating temperature of $40\mathrm{MeV}$, consistent with BBN constraints, and fix the initial PBH abundance to $\beta = 10^{-37}$. The shaded regions represent areas of DM overproduction. The blue, green, and magenta regions correspond to memory burden parameters $\kappa = 0$, $1$, and $2$, respectively. As expected, increasing $\kappa$ reduces the allowed parameter space due to the prolonged lifetime of PBHs. Additionally, the solid, dashed, and dotted lines correspond to $\lc = 0$, $\lcnr$, and $\lcr$, respectively. We see that accretion significantly impacts the viable parameter space of DM.

\subsection{Constraints from warm dark matter}

In this section, we aimed to satisfy the relic density of total DM of the Universe from the particle $j$, emitted from the evaporation of PBHs. In such a case, one needs to simultaneously check that the particles are cold enough so that they do not destroy structure formation. The DM particles should have a minimum mass in order to stream freely, as constrained from Ly-$\alpha$ power spectra. On the other hand, such light DM is significantly constrained due to the ultrarelativistic nature of the emitted particles. The lower bound on the mass of DM emitted from the evaporation of PBHs is given by~\cite{Auffinger:2020afu,Masina:2021zpu,Barman:2024iht}
\begin{align}
    m_j\geq 10^4 \langle E_{\rm eq}\rangle,
\end{align}
where $\langle E_{\rm eq}\rangle$ is the kinetic energy of the DM particles averaged over the temperature of PBHs. The average kinetic energy is further given by 
\begin{align}
    \langle E_{\rm eq}\rangle=
    \f{\aev}{\aeq}\langle E_{\rm ev}\rangle=
    \delta \Tbh^{\rm ev} \l(\f{g_{\rm s}^{\rm eq}}{g_{\rm s}^{\rm ev}}\r)^{1/3} \f{T_{\rm eq}}{\Tev}
\end{align}
where $g_{\rm s}^{\rm eq}=3.94$ and $g_{\rm s}^{\rm ev}=106.75$ are the relativistic degrees of freedom associated with entropy at the time of radiation-matter equality and at the time of evaporation. The quantity $T_{\rm eq}=0.75~{\rm eV}$ is the temperature at radiation-matter equality. The average kinetic energy at the time of evaporation is given in terms of the temperature of PBH at evaporation as $\langle E_{\rm ev}\rangle=
    \delta \Tbh^{\rm ev}$, where $\Tbh^{\rm ev}=\Mpl^2/(q\mac)$ and the parameter $\delta$ is taken to be $\delta\approx1.3$.
For the scenarios with PBH domination, the lower limit on the mass of emitted DM particles is given by
\begin{align}
    m_j\geq 0.02 \l(\f{g_{\rm s}^{\rm eq}}{g_{\rm s}^{\rm ev}}\r)^{1/3}
    \f{1}{\sqrt{2^\kappa(3+2\kappa)\epsilon}}
    \l(\f{q\mac}{\Mpl}\r)^{\kappa+1/2},
\end{align}
which is shown in the red lines in Fig.~\ref{fig:dm-bg}.

\subsection{Constraints from gravitationally produced dark matter}

While discussing the production of DM, one should also consider the production mechanism arising from gravitational interactions, as it is both natural and unavoidable. In this case, DM particles are produced through the exchange of a graviton~\cite{Garny:2015sjg,Bernal:2018qlk,Mambrini:2021zpp,Bernal:2021kaj,Barman:2021ugy,Haque:2021mab,Clery:2021bwz,Clery:2022wib,Ahmed:2022tfm,Chianese:2020khl,Maity:2024cpq,Franciolini:2026fdv}. 
In the gravitational production scenario, DM can be generated through two main channels: from the inflaton and from the thermal bath. However, in most cases, the production of both scalar and fermionic DM from the inflaton dominates over the production from the thermal bath, unless an extremely high temperature is assumed for the production of fermionic DM.
The present-day abundance of gravitationally produced DM can be written as~\cite{book,Maity:2024cpq}
\begin{align}
\frac{\Omega_j h^2}{0.12}= 1.3\times 10^9\,\frac{g_0}{g_{\ast}^{\rm re}}\,\frac{ n_s^\phi(\tre)}{\Tre^3}\,\frac{m_j}{\text{GeV}}\,,
\label{Eq:omegah2m}
\end{align}
where $g_0=3.91$ and $g_{\rm s}^{\rm re}\simeq 106.75$ denote the effective number of relativistic degrees of freedom for entropy evaluated at the present epoch and at the end of reheating, respectively. The quantity $n_s^\phi(\tre)$ represents the number density, where the subscripts $s=0$ and $s=1/2$ correspond to bosonic and fermionic DM, respectively. For a detailed discussion of this quantity, see~\cite{Maity:2024cpq}. 
In Fig.~\ref{fig:dm-bg}, we show the constraints or the upper bound arising from gravitationally produced fermionic DM, indicated by the purple lines.


\section{Stable PBHs as dark matter\label{sec:dm-stable}}

\begin{figure*}
    \centering
    \includegraphics[width=0.49\linewidth]{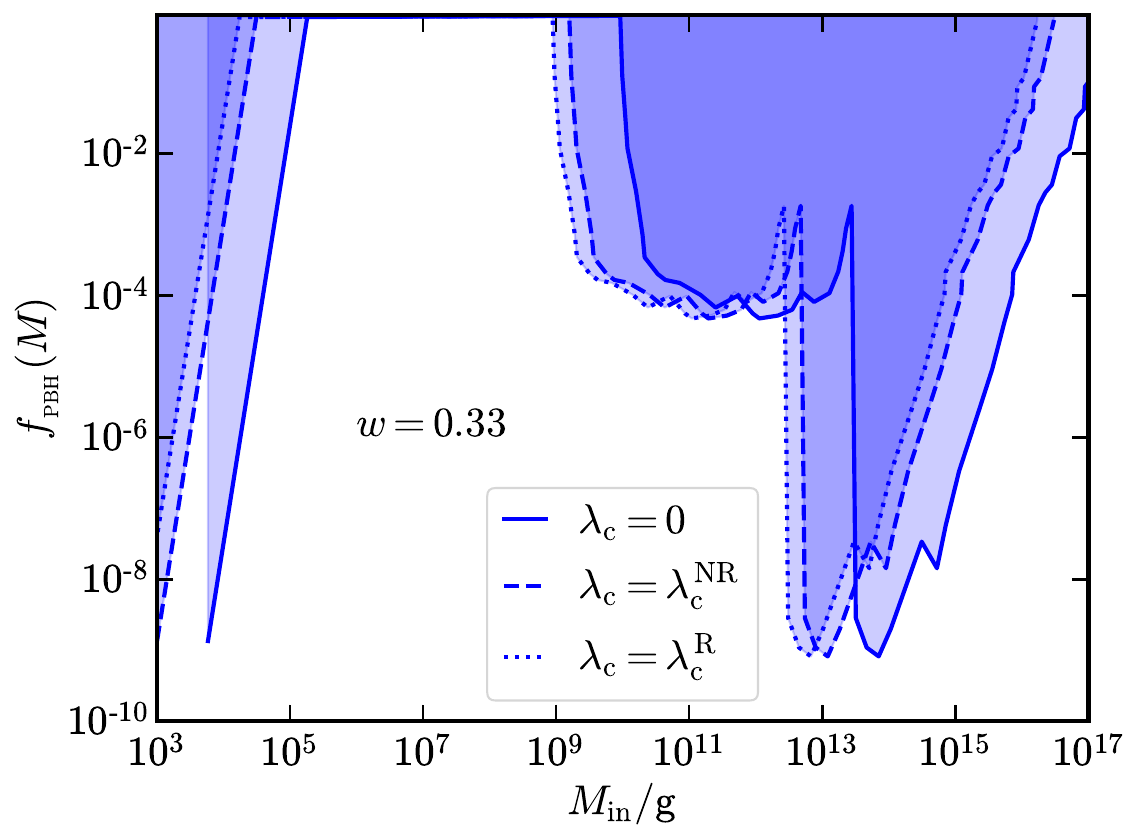}
    \includegraphics[width=0.49\linewidth]{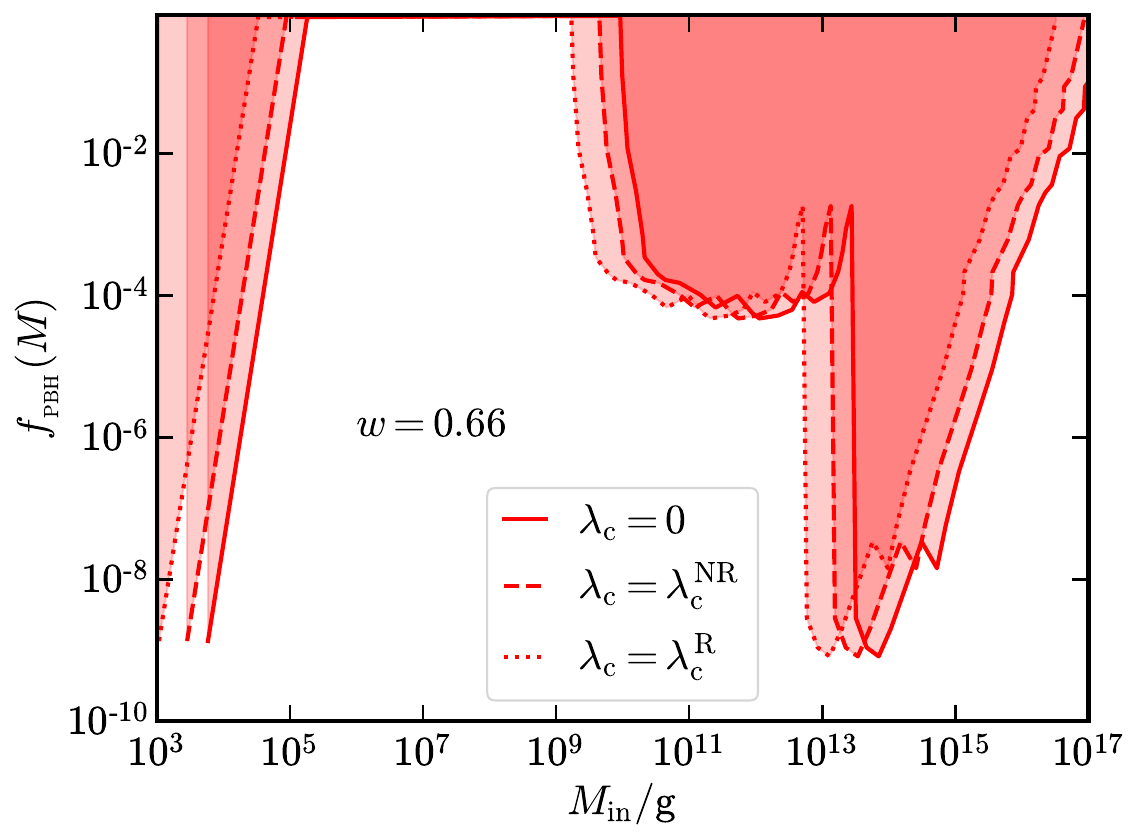}
    \caption{The constraints on the fraction of today’s DM energy density sourced by PBHs, denoted by $\fpbh$, is shown as a function of the PBH mass here. 
    Effects due to $w=1/3$ and $w=2/3$ are shown in the left and right panel respectively.
    The solid, dashed, and dotted lines correspond to the case of $\lc=0,~\lcnr, ~\lcr$ respectively.  The detailed descriptions of obtaining the constraints are given in~\cite{Ballesteros:2019exr,Carr:2020gox,Chen:2021ngo,Thoss:2024hsr,Chianese:2025wrk,Tan:2025vxp,Ambrosone:2026djo}.
    }
    \label{fig:fpbh}
\end{figure*}
As is now well understood, memory burden effects can stabilize lower-mass PBHs, allowing them to survive until the present Universe. In most cases,  for our choice of parameters,  the initial part of the Hawking evaporation of the PBHs has already occurred before BBN\footnote{ Note that we have taken the transitions to be sudden.}. Therefore, in principle, the total relic abundance of DM should include contributions from both the evaporated component and the remnant sector. However, the contribution from evaporation is typically subdominant compared to that of the remnant, unless the value of the parameter $q$ is very small (e.g., $q \lesssim 10^{-2}$); see~\cite{Haque:2024eyh} for details. Since we primarily focus on the case $q = 1/2$, the contribution from evaporation can be safely neglected. The dimensionless energy density parameter for the surviving PBHs is then given by
\begin{align}
    \obh=\f{\rhobh^0}{\rho_{\rm c}^0}=
    \f{5.62\times 10^9}{{\text{GeV}}}\f{g_0}{g_{\rm re}}\f{q\mac\nbh^{\rm re}}{\alpha \Tre^3},
\end{align}
where $g_0$ and $g_{\rm re }$ are the relativistic degrees if freedom at present day and at the time of reheating. The number density of PBHs at the time of reheating is given by 
\begin{align}
    \nbh^{\rm re}=\beta\f{\rhow^{\rm in}}{\Min}
    \l(\f{\ain}{\are}\r)^3,
\end{align}
with $\rhow^{\rm in}$ being the energy density of the background at the time of the formation of PBHs.
Next, the ratio of the scale factor during a $w$-dominated phase can be obtained as 
\begin{align}
    \l(\f{\ain}{\are}\r)^3&=\l(\f{\tin}{\tre}\r)^{\f{2}{1+w}}=
    \l(\f{\alpha}{48\pi^2}\r)^{\f{1}{1+w}} \gamma^{-\f{2}{1+w}}
    \nn\\&\quad
    \l(\f{\Tre}{\Mpl}\r)^{\f{4}{1+w}}
    \l(\f{\Min}{\Mpl}\r)^{\f{2}{1+w}}.
\end{align}
Finally, the dimensionless density parameter has the following form
\begin{align}
    \obh&= {5.16\times 10^{26}}\beta \f{q\mac}{\Min}
    \l(\f{48\pi^2}{\alpha}\r)^{\f{w}{1+w}}\gamma^{\f{2w}{1+e}}
    \nn\\&\quad
    \l(\f{\Tre}{\Mpl}\r)^{\f{1-3w}{1+w}}
    \l(\f{\Min}{\Mpl}\r)^{-\f{2w}{1+w}}.
\end{align}
An important point is that for $w=1/3$, $\obh$ becomes independent of the temperature of reheating. And if the PBHs form during an epoch of $w$-domination, $\obh$ behaves with the PBH mass as $\obh\propto \Min^{-2w/(1+w)}$. Another important quantity related to stable PBHs is $\fpbh$, which is defined as the fraction of the DM energy density today that comes from PBHs, and it is given by $\fpbh=\obh/0.12$.
As we have already seen, the effect of accretion substantially changes the lifetime of PBHs. Hence, the minimum value of PBH mass for which the PBH will survive till today also gets affected by the accretion.   
In Fig.~\ref{fig:fpbh}, we have shown
such effects of accretion on the constraints on $\fpbh$.
In the left panel of Fig.~\ref{fig:fpbh}, we have chosen $w=1/3$ and in the right panel, we have chosen $w=2/3$. The solid, dashed, and dotted lines correspond to the case of $\lc=0,~\lcnr, ~\lcr$ respectively.
Firstly, let us consider the region of $\Min\geq 10^{10}$ g. The constraints above this range of mass are not affected by the memory burden parameter, as the semiclassical evolution of PBHs still goes on for the PBH mass. But while the accretion is involved, a PBH with a lower initial mass will now also remain in the semiclassical process due to its mass gain from accretion. On the other hand, for PBHs with $\Min\leq 10^{5}$ g, the constraints on $\fpbh$ have effects due to the evaporation of PBHs. With a similar argument as before, it is easy to see that now a lower mass PBH will contribute to these constraints due to accretion.

In the outline of the constraints on $\fpbh$ shown in Fig.~\ref{fig:fpbh}, the total constraint arises from four different sources. Below, we briefly describe how each of these individual constraints is obtained. For detailed discussions about the constraints, see~\cite{Ballesteros:2019exr,Carr:2020gox,Chen:2021ngo,Auffinger:2022khh,Thoss:2024hsr,Chianese:2025wrk,Ambrosone:2026djo}.
The emission rate of the particles produced through PBH evaporation is given by 
\begin{align}
    \f{\d^2N_i}{\d E\d t}=\f{g_i}{2\pi}\f{\cF(E,M)}{\exp({E/\Tbh})-(-1)^{2s_i}},
\end{align}
with $g_i$ and $s_i$ be the internal degrees of freedom and the spin of the emitted particles, respectively. The quantity $\cF(E,M)$ is known as graybody factor.
\\ 
\\
$\bullet$ {\it Galactic $\gamma$-ray background:}
    As already discussed, PBHs can emit photons via evaporation. If a significant amount of DM in the Milky Way DM halo comes from PBHs, then the photon emitted via their evaporation shall be observed. The measured flux is given by~\cite{Ballesteros:2019exr,Carr:2020gox,Chen:2021ngo,Thoss:2024hsr,Tan:2025vxp,Chianese:2025wrk,Ambrosone:2026djo} 
    \begin{align}
        \Phi_{\rm PBH}^{\rm gal} &= 
        \f{\fpbh}{4\pi M\,\Delta\Omega} \f{\d^2N_\gamma}{\d E\d t} 
        \int\d\Omega\int\d s
        \,\rho_{_\mathrm{DM}}(r)
        \e^{-\tau(E_\gamma,b,l)}
    \end{align}
where ${\d^2N_\gamma}/{\d E\d t}$ is the $\gamma$-ray emission rate, $\Omega$ is the solid angle, $r$ is the galactocentric radial coordinate, $b$ and $l$ are the galactocentric lattitude and longitude, respectively.
$\gamma$-rays further interact with background photons and create electron-positron pairs, where $\tau(E_\gamma,b,l)$ indicates the corresponding optical depth and $\exp\l[-\tau(E_\gamma,b,l)\r]$ accounts for the attenuation.
The constraint on $\fpbh$ is obtained by 
\begin{align}
    \int_{E_{\rm min}}^{E_{\rm max}}
    \d E \,\Phi_{\rm PBH}^{\rm gal}
    \leq \Phi^{\rm gal}({E_{\rm max}}-{E_{\rm min}}),
\end{align}
where $\Phi_{\rm PBH}^{\rm gal}$ is the measured value of the $\gamma$-ray flux in the energy window $[{E_{\rm min}},{E_{\rm max}}]$.
\\
\\
$\bullet$ {\it Extragalactic $\gamma$-ray background:} 
The extragalactic $\gamma$-ray constraint arises from the cumulative emission of photons produced by PBH evaporation from the time of recombination until today. The measured flux in such cases is given by~\cite{Ballesteros:2019exr,Carr:2020gox,Chen:2021ngo,Thoss:2024hsr,Chianese:2025wrk,Tan:2025vxp,Ambrosone:2026djo} 
\begin{align}
    \Phi_{\rm PBH}^{\rm egal} &= 
        \f{\fpbh\,\Omega_{\rm DM}\,\rho_{\rm c}}{4\pi  M} 
        \int_0^{z_{\rm rec}} \f{\d z}{H(z)}\nn\\
        &\quad\times \l[(1+z)^{-1} \f{\d E_i}{\d E}  \f{\d^2N_\gamma}{\d E\d t}\Big\vert_{E=E_i}\r]
\end{align}
where $z$ is the redshift and the subscript 'rec' refers to the time of recombination, $E_i$ is the energy od the photon at the time of recombination. The quantity $\Omega_{\rm DM}=0.12$ is the density parameter of DM and $\rho_{\rm c}=4.79\times10^{-6}\,{\rm GeV/cm^2}$ is the critical energy density of the Universe. 
\\
\\
$\bullet$ {\it CMB anisotropies:} 
During recombination, electrons and protons combine to form neutral hydrogen, which allows photons to decouple and produce CMB. However, if PBHs evaporate after this epoch, the emitted high-energy particles and photons deposit energy into the intergalactic medium. This injected energy can ionize neutral hydrogen and helium.
The energy injection rate per unit volume is given by~\cite{Poulin:2016anj,Stocker:2018avm,Acharya:2019xla,Carr:2020gox,Thoss:2024hsr}
\begin{align}
    \f{\d^2E}{\d V\d t}  =\f{\fpbh\,\Omega_{\rm DM}\,\rho_{\rm c}}{ M} h_\alpha(z)(1+z)^3 \l\vert \f{\d M}{\d t} \r\vert,
\end{align}
where, as not all injected energy affects the plasma, a deposition efficiency factor $ h_\alpha(z)$ is multiplied. CMB observations limit the amount of energy injected at a given redshift $z$.
\\
\\
$\bullet$ {\it Big bang nucleosynthesis:}
PBHs that evaporated during BBN, inject high-energy neutrinos and antineutrinos that change the neutron-to-proton ratio at the onset of BBN. Evaporation of PBHs at the time increases the baryon-to-entropy ratio, which results in overproduction of ${}^4 \rm He$.
The constraints on the abundance is given by~\cite{Carr:2020gox,Thoss:2024hsr,Kawasaki:2017bqm}
\begin{align}
    \beta=5.4\times 10^{21} \l(\f{t_{\rm ev}}{1\,\rm sec}\r)^{1/2}\f{\nbh}{s},
\end{align}
where $\nbh$ is the number density of PBHs and $s$ is the entropy density.
\\
\\
As the accretion changes the initial mass of the PBHs, it affects the evolution of the PBH mass along with the evaporation time. Thus, in the constraints on $\fpbh$, we observe the shift in the formation mass.


\section{Dark radiation from the evaporation of PBHs \label{sec:DR-evap}}

\begin{figure*}
    \centering
    \includegraphics[width=0.49\linewidth]{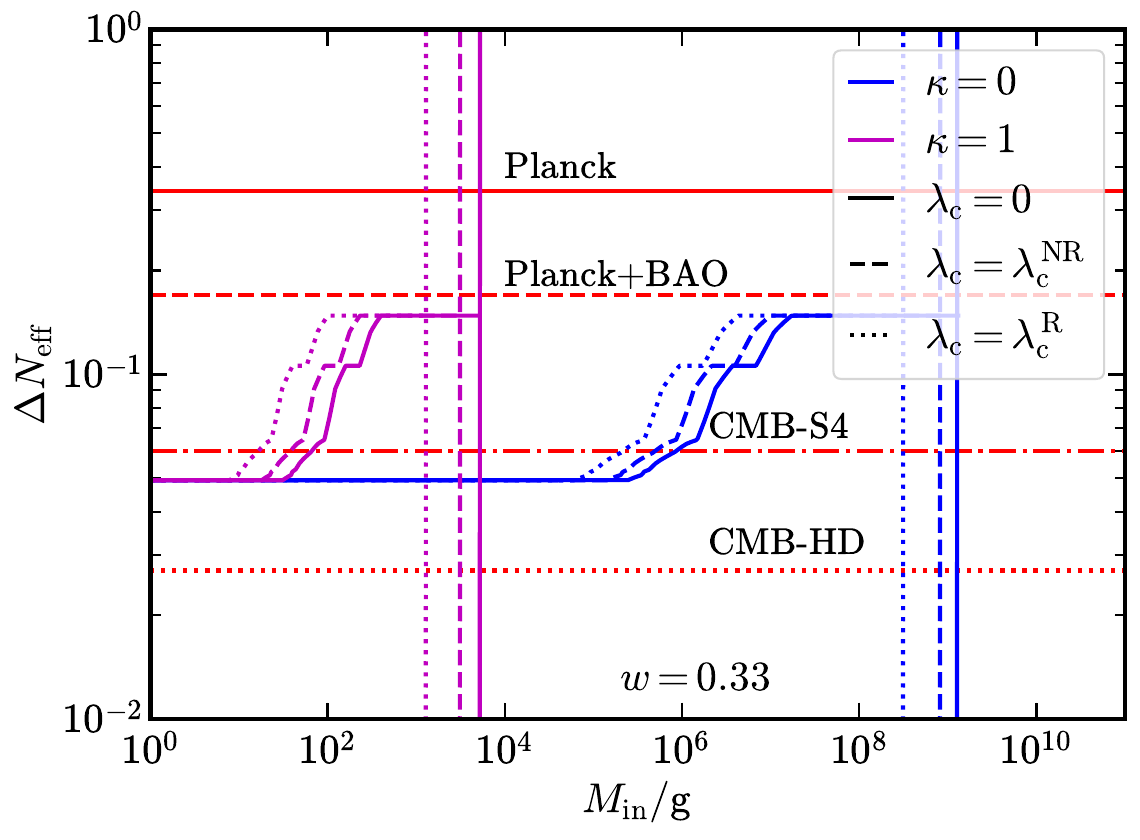}
    \includegraphics[width=0.49\linewidth]{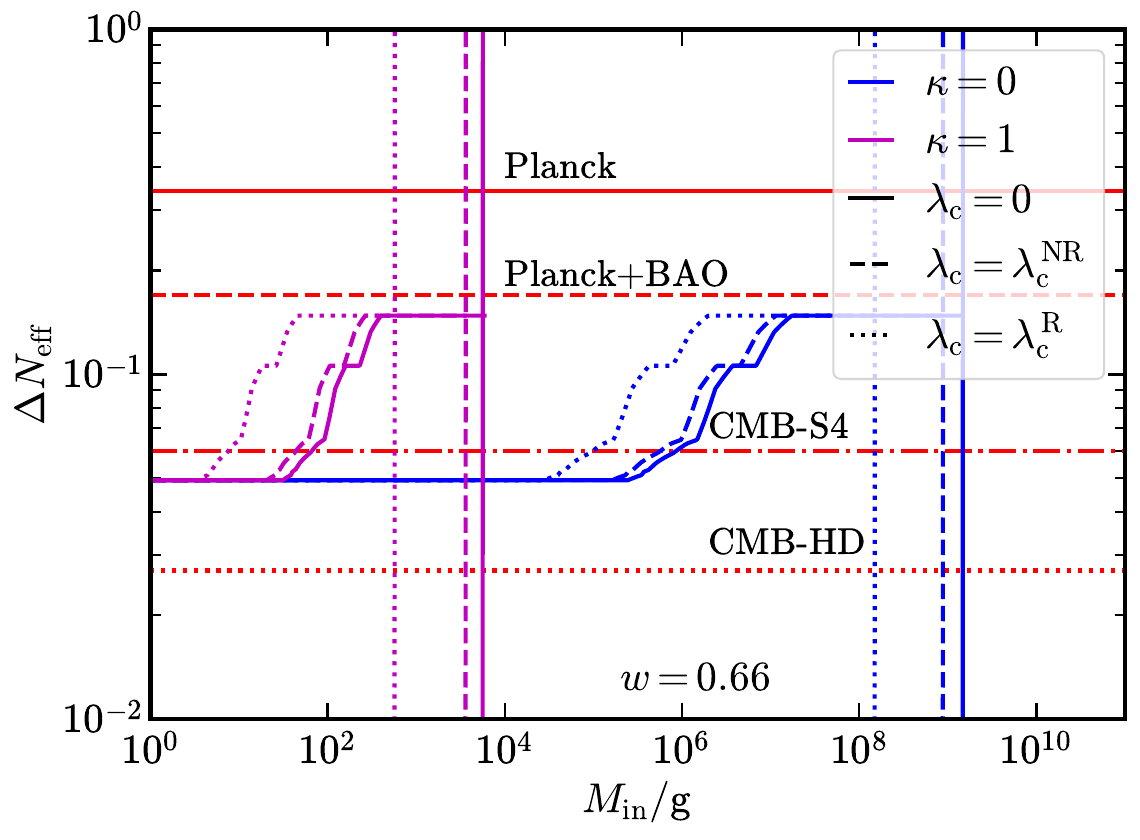}
    \caption{The contribution to $\dneff$ from spin-0 bosonic particles generated from the evaporation of PBHs is shown as a function of the initial mass of PBHs, for two different backgrounds where the PBHs have formed, i.e. $w=1/3$ (left panel) and $w=2/3$ (right panel), respectively.
The red solid, dashed, dot-dashed, and dotted lines correspond to the bounds on $\dneff$ given by observations from Planck, Planck+BAO, CMB-S4, and CMB-HD. The blue and magenta lines correspond to two different choices of the memory burden parameters, $\kappa=0,\,q=1$ and $\kappa=1,\,q=1/2$, respectively.
The vertical lines represent the maximum limit on the formation mass that ensures complete evaporation of PBHs before BBN. As usual, the cases of $\lc=0$, $\lcnr$, and $\lcr$ are shown with solid, dashed, and dotted lines, respectively.
    }
    \label{fig:dneff1}
\end{figure*}
As the PBHs evaporate, along with SM particles, other species such as DM, right-handed neutrinos, and DR can also be produced. DR refers to light, feebly interacting particles that remain relativistic during early cosmic epochs and thus contribute to the total radiation energy density at the time of BBN or even during recombination. The presence of such DR is commonly characterized by its contribution to the effective number of relativistic degrees of freedom, denoted by the quantity $\dneff$, which is defined as~\cite{Masina:2021zpu}
\begin{align}
    \dneff=\f{\rho_{_\mathrm{DR}}^{\rm eq}}{\rhor^{\rm eq}}
    \l[N_\nu + \f{8}{7}\l(\f{11}{4}\r)^{4/3}\r],
    \label{eq:dneff1}
\end{align}
where $\rho_{_\mathrm{DR}}^{\rm eq}$ and $\rhor^{\rm eq}$ are the energy density of DR and radiation at the time of radiation-matter equality, and $N_{\nu}=3.045$ is the effective number of neutrino species~\cite{Dodelson:1992km,Hannestad:1995rs,Dolgov:1997mb,Mangano:2005cc,deSalas:2016ztq,EscuderoAbenza:2020cmq,Akita:2020szl,Froustey:2020mcq,Bennett:2020zkv}. Using the conservation of entropy, as defined before, the ratio of the energy densities given by 
\begin{align}
    \f{\rho_{_\mathrm{DR}}^{\rm eq}}{\rhor^{\rm eq}}=
    \f{g_{\rm ev}}{g_{\rm eq}}\l(\f{g_{\rm s}^{\rm eq}}{g_{\rm s}^{\rm ev}}\r)^{4/3}
    \f{\rho_{_\mathrm{DR}}^{\rm ev}}{\rhor^{\rm ev}},
\end{align}
where $g_{\rm ev}$ and $g_{\rm eq}$ are the relativistic degrees of freedom and $g_{\rm s}^{\rm ev}$ and $g_{\rm s}^{\rm eq}$ are relativistic degrees of freedom associated with entropy at the time of evaporation and at radiation-matter equality, respectively. 
For the scenario with PBH domination, we get ${\rho_{_\mathrm{DR}}^{\rm eq}}/{\rhor^{\rm eq}}={g_{_{\rm DR}}^{_{\rm H}}}/{g_{_{\rm H}}}$, and hence Eq.~\eqref{eq:dneff1} reads~\cite{Masina:2021zpu}
 \begin{align}
     \dneff=
    \f{g_{\rm ev}}{g_{\rm eq}}\l(\f{g_{\rm s}^{\rm eq}}{g_{\rm s}^{\rm ev}}\r)^{4/3} \f{g_{_{\rm DR}}^{_{\rm H}}}{g_{_{\rm H}}}
    \l[N_\nu + \f{8}{7}\l(\f{11}{4}\r)^{4/3}\r],
 \end{align}
 where $g_{_{\rm DR}}^{_{\rm H}}=4,\,2,\,1.82,\,1.23,\,0.05$ for Dirac and Weyl fermion, scalar and vector boson, and massless graviton, respectively. 
Observation of CMB currently restricts $\dneff$ to $\dneff=0.34$~\cite{Planck:2018jri}, whereas combined with data of baryon acoustic oscillation (BAO), the constraint improves to $N_{\rm eff}=2.99\pm0.17$.
Further, forthcoming CMB missions such as CMB-S4~\cite{Abazajian:2019eic}, CMB-HD~\cite{CMB-HD:2022bsz} are expected to measure $\dneff$ with more precisions of $\dneff=0.06$, and $\dneff=0.027$, respectively. 
In Fig.~\ref{fig:dneff1}, we have shown the contribution to $\dneff$ from the spin-0 bosonic components for the case of PBH domination. In the case of $\dneff$, the detailed behavior of $g_{\rm s}^{\rm ev}$ with $\Tev$ is required, where for the case of PBH domination, the temperature at evaporation reads
$\Tev=\l[(4/(3\alpha))\Mpl^4 2^{2\kappa}(3+2\kappa)^2\epsilon^2(\Mpl/(q\mac))^{6+4\kappa}\r]^{1/4}$.
The blue and magenta lines correspond to $\kappa=0,\,q=1$ and $\kappa=1,\,q=1/2$, respectively, while the solid, dashed, and dotted lines correspond to $\lc=0$, $\lcnr$, and $\lcr$.
In the figure, we have illustrated the case for a single component with $g_{{\rm DR}}^{{\rm H}}=1.82$, as our aim is to show the effects of $\kappa$ and $w$. Contributions from other sectors with different values of $g_{{\rm DR}}^{{\rm H}}$ will have similar qualitative effects. It can be shown that the upcoming observations from CMB-S4 or CMB-HD might be able to rule out all components except the massless graviton as contributors to DR.




\section{Conclusions \label{sec:conclusion}}

In this work, we have examined the effects of relativistic accretion and burdened evaporation on the evolution of PBHs. The quantum memory burden effect, which induces a backreaction on the process of evaporation, causes PBHs to persist for a longer duration. On the other hand, accretion of relativistic particles can lead to a substantial growth in the initial mass of PBHs, allowing them to survive even longer (see Figs.~\ref{fig:m-acc-evap} and~\ref{fig:k-min-1}). Our primary aim was to investigate the joint effects of these two processes in the context of various phenomena, which we summarize below.

We considered PBHs with an initial mass range such that they are expected to have evaporated before BBN. Throughout this work, we have explored two choices for the background EoS, namely $w=1/3$ and a stiffer EoS $w=2/3$, as toy examples. We focused on scenarios where the evaporation of PBHs leads to the production of DM particles. For the evaporation process, we analyzed two distinct cases characterized by the critical value of the initial abundance of PBHs, $\beta$. When $\beta > \betacr$, the energy density of PBHs dominates over the background energy density before evaporation, whereas for $\beta < \betacr$, the energy density of PBHs remains subdominant.
For the case $\beta < \betacr$, we further considered a scenario where the background is governed by $w > 1/3$, and reheating is achieved when the radiation generated from the evaporation of PBHs eventually dominates the background energy density. We demonstrated that increasing the memory burden parameter $\kappa$ significantly {changes} the available parameter space in the $m_j - \Min$ plane for the emitted mass of DM and the initial mass  of PBHs (see Figs.~\ref{fig:dm-bg} and~\ref{fig:dm-bl}). Including the effects of accretion further shrinks this parameter space, with relativistic accretion leading to particularly pronounced reductions. Additionally, we provided constraints on the $m_j - \Min$ parameter space arising from warm DM.

We further considered the mass ranges of PBHs and the memory burden parameters for which PBHs can survive until the present epoch. We calculated the fraction of the total DM given by PBHs, denoted by $\fpbh$, and examined the impacts of both accretion and burdened evaporation on $\fpbh$. We demonstrated that while the memory burden effect opens up a new window of $\fpbh$ for smaller masses of PBHs, accretion shifts both the mass of PBHs and the corresponding value of $\fpbh$s to higher values for a given initial mass. Building on this, we explored how accretion influences the constraints from various mass-dependent observational limits on $\fpbh$ (see Fig.~\ref{fig:fpbh}).

Lastly, we considered the emission of DR from the evaporation of PBHs and examined their effects on $\dneff$, (see Fig.~\ref{fig:dneff1}), especially when the effects of accretion and memory burden are considered. 

{
This work has briefly discussed the combined effects of relativistic accretion and burdened evaporation. The ideas presented here can be extended in several directions. For instance, one may consider the case of Kerr black holes, where spherical symmetry is broken, and investigate accretion together with evaporation in that context. 
Moreover, in this work, we have restricted our analysis to the case in which the transition from semiclassical evaporation to the memory-burdened phase is assumed to be sudden. It would also be worthwhile to explore the scenario of a gradual transition to the memory-burdened regime~\cite{Dvali:2025ktz,Montefalcone:2025akm} and examine the resulting implications.
}


\section*{Acknowledgments}
SM thanks the Indian Institute of Science Education
and Research (IISER), Pune, for support through a postdoctoral fellowship.

\bibliographystyle{apsrev4-1}
\bibliography{references}

\end{document}